\documentclass[%
 preprint, 
 amsmath,amssymb,
 aps, physrev,
]{revtex4-2}

\usepackage{graphicx}
\usepackage{dcolumn}
\usepackage{bm}
\usepackage{physics}
\usepackage{xcolor}

\begin{document}

\title{Effect of high odrer dispersion on soliton interaction in coupled Kerr cavities }%

\author{A. G. Vladimirov}
\email{corresponding author: an\_drei@list.ru}
\affiliation{National Research University Higher School of Economics, 25/12 Bolshaya Pecherskaya Ulitsa, 603155 Nizhny Novgorod, Russia}

\date{\today}


\begin{abstract}
We investigate the interaction of dissipative Kerr solitons in two nearly identical coherently driven, coupled microresonators in the presence of third-order dispersion (TOD). Using an asymptotic reduction of the coupled Lugiato--Lefever equations, we derive effective interaction equations governing the soliton separation and collective drift for weak intercavity coupling. TOD breaks the parity symmetry of the soliton profile and qualitatively modifies the effective interaction, leading to asymmetric interaction forces and multiple stable and unstable soliton clusters. The existence and stability of these clusters are controlled by the relative phase of the driving fields and the group-velocity mismatch between the resonators. In particular, an appropriate value of the group-velocity mismatch can give rise to multiple stable bound states, whereas deviations from this value trigger successive saddle-node bifurcations, ultimately leading to periodic soliton collisions. The analytical predictions are confirmed by numerical simulations of the full coupled system.
\end{abstract}

\maketitle

\section{Introduction}

Kerr microresonators have emerged as a versatile platform for nonlinear photonics, enabling optical frequency comb generation through dissipative soliton formation \cite{herr2014temporal,kippenberg2018dissipative}. Dissipative Kerr cavity solitons arise from the interplay of Kerr nonlinearity, dispersion, coherent driving, and dissipation, maintaining their temporal profiles as they circulate within the resonator \cite{herr2016dissipative}. While most studies of Kerr soliton dynamics focus on a single continuously pumped cavity, configurations involving coupled microresonators  remain comparatively less explored \cite{xue2019super,helgason2021dissipative,komagata2021dissipative,tikan2022protected,yuan2023soliton,tusnin2023nonlinear,dolinina2026soliton,kolesnikova2026hybridization}. Yet systems of multiple interacting Kerr cavities offer a promising route toward scalable on-chip photonic architectures, enabling tunable filtering, signal routing, and all-optical processing. 

An important ingredient underlying the rich dynamics in optical microcavities is the weak interaction between Kerr solitons. Arising from the overlap of exponentially decaying tails or through long-range perturbative effects, these interactions govern the formation of ordered structures and collective behavior in dissipative nonlinear media \cite{gorshkov1981interactions,karpman1981perturbational,Malomed1991,Akhmediev1997,vladimirov2001stable,akhmediev2005dissipative,Descalzi2011,turaev2012long,parra2014dynamics,vladimirov2023temporal}. They are responsible for the emergence of soliton clusters---often termed soliton molecules or crystals---as well as self-organization and symmetry-breaking effects. Theoretically, such phenomena in Kerr microcavities are commonly described by the Lugiato--Lefever equation (LLE) \cite{Lugiato1987} and its extensions. Soliton clusters have been observed experimentally in both fibre-loop oscillators \cite{Tang2005,Runge2015} and integrated microresonators \cite{herr2014temporal,kippenberg2018dissipative}, and are of interest for frequency comb generation, optical storage, and ultrafast signal manipulation.

Despite significant progress, the weak-interaction regime remains challenging to analyze because the interaction forces become exponentially small with increasing pulse separation. In this limit, even minor perturbations—for instance, periodic modulation \cite{turaev2012long} or high-order dispersive effects \cite{parra2014third,milian2015solitons,brasch2016photonic,cherenkov2017dissipative,parra2017interaction,vladimirov2018effect,weng2019polychromatic,weng2019polychromatic,vladimirov2021dissipative,parra2022quartic,anderson2022zero,todd2023dynamics,bahloul2026controlling}—can have pronounced influence. In particular, although typically small compared with second-order dispersion,  third order dispersion (TOD)  can exert a disproportionate influence on soliton tails because these tails are governed by the far-detuned spectral components of the dispersion relation, where the cubic dependence of the TOD term becomes increasingly significant. Consequently, even modest TOD can reshape the exponentially decaying wings, induce oscillatory dispersive-wave contributions, also known as Cherenkov radiation, and alter both the phase and amplitude of the soliton tails \cite{akhmediev1995cherenkov,cherenkov2017dissipative}. Because interactions between well-separated solitons are mediated by their tails, these modifications can substantially alter the weak inter-pulse forces, enabling bound states that would not exist without TOD. More generally, TOD breaks temporal symmetry, introduces asymmetry into soliton profiles, induces spectral recoil and temporal drift, and modifies interaction potentials, thereby influencing the formation, stability, and long-range dynamics of soliton bound states.

In a recent study, the interaction dynamics of dissipative solitons in two coherently pumped, coupled Kerr microresonators were examined using coupled LLEs \cite{dolinina2026soliton}. That analysis, which combines asymptotic techniques with numerical methods, revealed three distinct types of soliton clusters, distinguished by the inter-soliton separations within each cluster: ``bunching'' states, ``anti-bunching'' configurations, and bound-state clusters. The stability of these collective regimes was found to depend primarily on the relative phase of the two injection fields, while the equilibrium separation between solitons within a cluster could be effectively controlled by adjusting the group-velocity mismatch between the cavities.

In the present work, we focus specifically on the effect of TOD on the soliton interaction in this coupled-cavity system. To investigate this effect, we perform an asymptotic reduction of the coupled LLEs augmented with a TOD term, deriving an effective interaction equations that depend on the TOD coefficient. The analytical predictions are then validated and extended through direct numerical simulations of the full system. Our results demonstrate that TOD qualitatively modifies the interaction landscape, giving rise to new bound states, inducing systematic drift of two-soliton clusters, and shifting the parameter ranges forbound-state configurations. We show that the number of possible soliton clusters can be controlled by appropriately tuning the group-velocity mismatch between the two cavities. These findings establish higher-order dispersion as an additional control knob for manipulating soliton arrangements in coupled photonic structures and provide further insight into the rich phenomenology of dissipative soliton complexes in microresonator platforms.

\section{Model equations}

\begin{figure}[thb]
\centering
\includegraphics[width=0.9\linewidth]{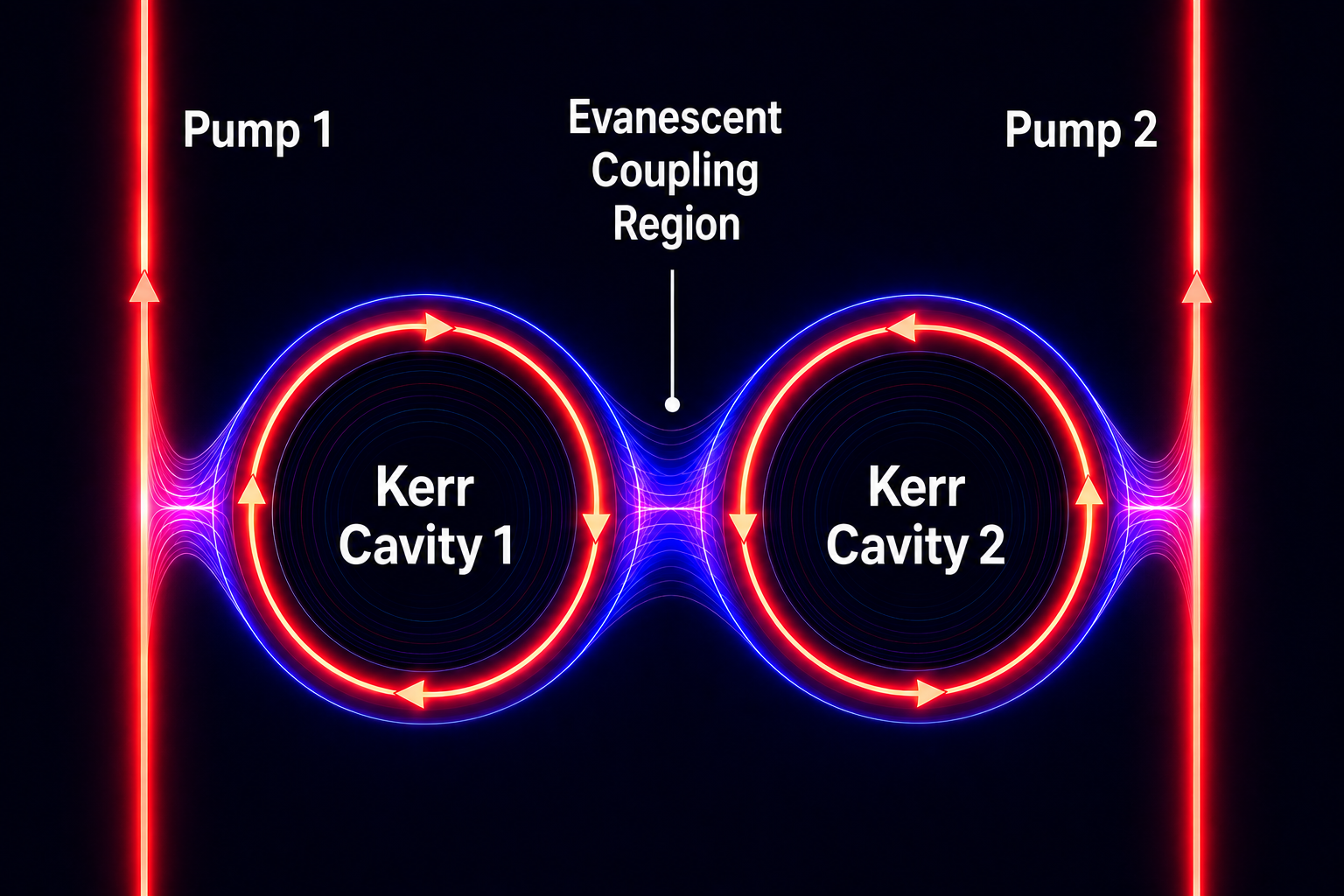}
\caption{Schematic view of the two coupled ring resonators driven by coherent pump beams. Arrows mark the light propagation direction. }
\label{fig:scheme}
\end{figure}

We consider a dual-ring resonator system, where two cavities are driven by phase-locked external fields and interact via evanescent coupling. A schematic illustration of the setup is provided in Fig.~\ref{fig:scheme}. Within the mean-field framework, the evolution of the intracavity fields is governed by a pair of coupled LLEs \cite{dolinina2026soliton}:

\begin{gather}
    \partial_t A_1 = -\dfrac{V}{2} \partial_\xi A_1 + i b_1 \partial_{\xi\xi} A_1 + d_1 \partial_{\xi\xi\xi} A_1 + i |A_1|^2 A_1 - (k_1 + i\theta_1) A_1 + i\kappa A_2 + i\eta_1 e^{i\delta/2}, \label{eq:LLE_coupled1}\\
    \partial_t A_2 = \dfrac{V}{2} \partial_\xi A_2 + i b_2 \partial_{\xi\xi} A_2 + d_2 \partial_{\xi\xi\xi} A_2 + i |A_2|^2 A_2 - (k_2 + i\theta_2) A_2 + i\kappa A_1 + i\eta_2 e^{-i\delta/2}.
    \label{eq:LLE_coupled2}
\end{gather}
Here, $A_{1,2}(t,\xi)$ denote the slowly varying envelopes of the optical pulses, where $t$ represents the slow-time variable tracking successive round trips and $\xi$ is the fast retarded time within the cavity. The parameter $\kappa$ characterizes the strength of the intercavity coupling, while $\eta_{1,2}$ denote the pump amplitudes. The phase offset between the two driving fields is represented by $\delta$. The coefficients $b_{1,2}$ and $d_{1,2}$ describe second-order group-velocity dispersion and TOD, respectively, with the latter introducing asymmetry into the dispersion profile and becoming particularly relevant for broadband pulses or operation near the zero-dispersion wavelength. The parameters $k_{1,2}$ quantify linear cavity losses, while $\theta_{1,2}$ denote the cavity detunings. Finally, $V$ accounts for a slight mismatch in the first-order dispersion between the two resonators, reflecting differences in their optical round-trip times.

The system (\ref{eq:LLE_coupled1}) and (\ref{eq:LLE_coupled2}) can be derived from a set of neutral delay differential equations ~\cite{vladimirov2023neutral} describing two mutually coupled cavities in the mean-field limit~\cite{dolinina2026soliton}. This reduction presumes that the field envelopes obey periodic boundary conditions $A_{1,2}(t,0)=A_{1,2}(t,T)$, with $T$ being the round-trip time. The pump fields are assumed to originate from a single laser source split into two paths, which guarantees mutual coherence and equal optical frequencies. The global phase $\phi$ common to both injection terms can be removed by applying the gauge transformation $A_{1,2} \to A_{1,2}\exp(i\phi)$, which also allows one to eliminate the factor $i$ multiplying the driving terms. We focus on the regime where the two cavities are nearly identical and the pump powers are comparable. Their magnitudes are selected such that each resonator, when isolated, supports a stable dissipative soliton. We deliberately exclude injection levels beyond the Andronov--Hopf instability threshold, since in that regime soliton interactions exhibit richer and more intricate behavior, as illustrated, for example, in the case of two solitons coexisting in a single Kerr cavity above this threshold~\cite{turaev2012long}.

\section{Soliton interaction equations}\label{sec:asympt}

We now develop an asymptotic description of two-soliton interaction in the coupled-cavity system described by Eqs.~(\ref{eq:LLE_coupled1}) and (\ref{eq:LLE_coupled2}), under the assumption of weak intercavity coupling. When $\kappa$ is small, the phase difference between solitons residing in different cavities remains close to $\delta$, the phase offset of the two driving fields. This phase difference determines whether the solitons attract or repel each other and governs the formation of bound complexes. To proceed, we set $\kappa = \epsilon K$ and $V = \epsilon u$, with $\epsilon \ll 1$, and assume the two resonators are nearly identical. Their parameters are expanded as $k_{1,2}=k+r_{1,2}$, $\theta_{1,2}=\theta+\Theta_{1,2}$, $b_{1,2}=b+B_{1,2}$, $d_{1,2}=d+D_{1,2}$, and $\eta_{1,2}=\eta+I_{1,2}$, where all correction terms are of order $\epsilon$. We look for solutions of the form
\begin{equation}
A_{1,2}(t,\xi)=i\left[A^{(0)}_{1,2}(\tau,\xi)+\epsilon A^{(1)}_{1,2}(t,\xi)+{\cal O}(\epsilon^2)\right]e^{\pm i\delta/2+{\cal O}(\epsilon^2)},\label{eq:ansatz}
\end{equation}
where $A^{(0)}_{1,2}(\tau,\xi)=A^{(0)}[\xi-\xi_{1,2}(\tau)]$ and $A^{(0)}(\xi)$ is the soliton solution of Eq.~(\ref{eq:LLE_coupled1}) [Eq.~(\ref{eq:LLE_coupled2})] with $\delta=-\pi$ ($\delta=\pi$) and $\epsilon=0$. The quantities $\xi_{1,2}(\tau)$ represent the slowly drifting positions of the two solitons, with $\tau=\epsilon t$ denoting the slow time scale. Substituting this expansion into the governing equations and extracting terms of order $\epsilon^0$ yields two uncoupled LLEs with TOD, which are identically satisfied by the unperturbed solitons. At order $\epsilon^1$, we obtain
\begin{equation}
{\cal L}_{1,2}A^{(1)}_{1,2}=\psi_{1,2}\left(\partial_\tau \xi_{1,2}\mp \frac{u}{2}\right)+iK A^{(0)}_{2,1}\exp(\mp i\delta)-f_{1,2},\label{eq:1storder}
\end{equation}
where the linear operator
\[{\cal L}_{1,2}A^{(1)}_{1,2}= \left(i b \partial_{\xi\xi} + d \partial_{\xi\xi\xi} + i |A_{1,2}^{(0)}|^2-1 - i\theta\right)A^{(1)}_{1,2}  + 2 i A_{1,2}^{(0)2} A^{(1)*}_{1,2}\]
is obtained by linearizing the single-cavity LLE (including TOD) around $A^{(0)}_{1,2}$. The source term is given by $f_{1,2}=-(r_{1,2}+i\Theta_{1,2})A_{1,2}^{(0)}+iB_{1,2}\partial_{\xi\xi}A_{1,2}^{(0)}+D_{1,2}\partial_{\xi\xi\xi}A_{1,2}^{(0)}+i I_{1,2} e^{\pm i\delta/2}$, and $\psi_{1,2}=\psi(\xi-\xi_{1,2})=\partial_\xi A^{(0)}_{1,2}$ are the neutral eigenmodes associated with the continuous translational invariance of the uncoupled system, ${\cal L}_{1,2}\psi_{1,2}=0$. Unlike the TOD-free case, the inclusion of TOD breaks the parity symmetry of the soliton profile, so $A^{(0)}(\xi)$ is no longer an even function and the adjoint neutral mode lacks definite parity. 

\begin{figure}[thb]
\centering
\includegraphics[width=0.9\linewidth]{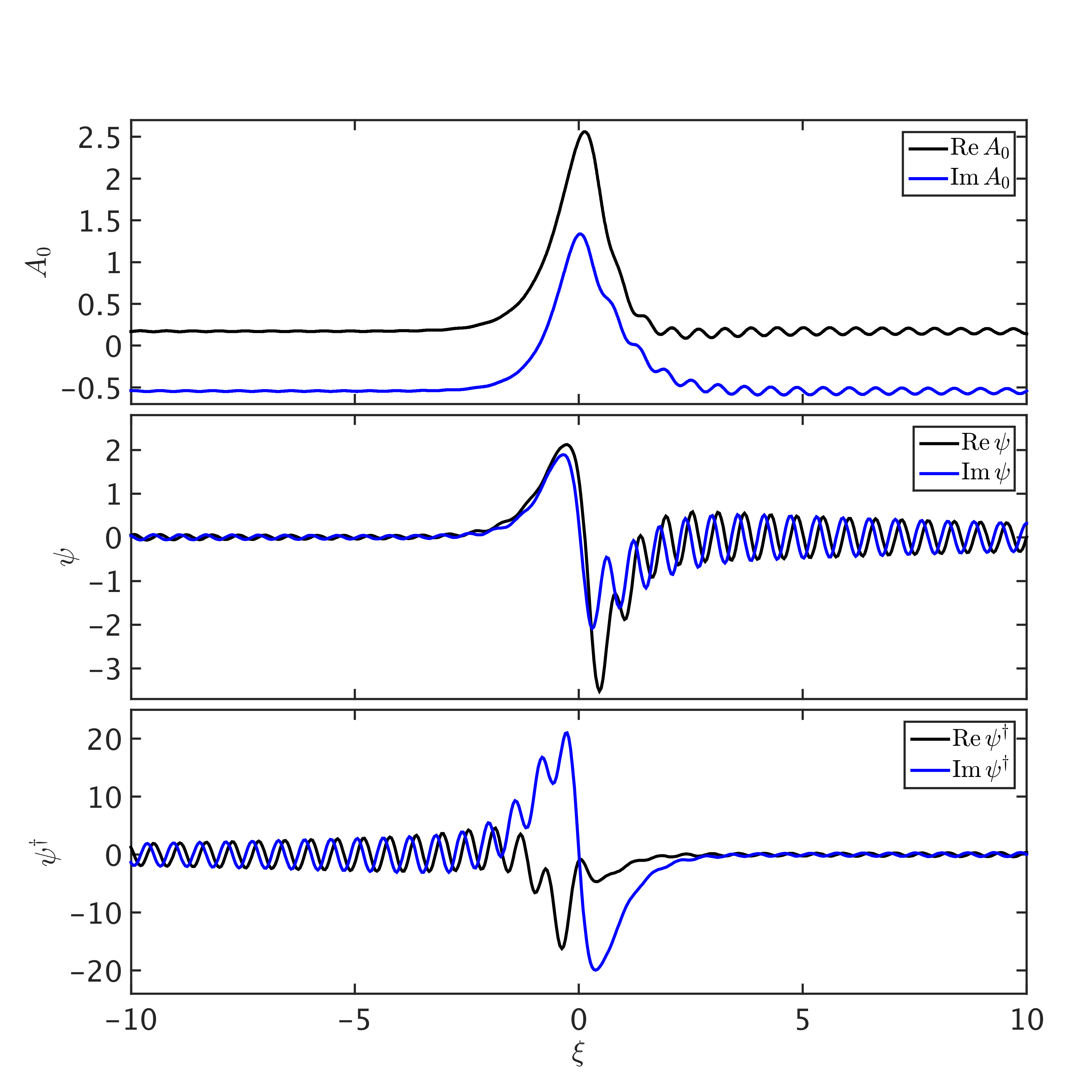}
\caption{Real (black) and imaginary (blue) parts of the soliton solution (top), the neutral mode (middle), and the adjoint neutral mode (bottom) for the uncoupled LLE with $\epsilon=0$. The parameters are $\theta=3.5$, $k=1$, $\eta=1.9$, $\delta=-\pi$,  $b=1$, $d=0.1$, and $T=40.0$. For improved visualization, the solutions are plotted over an interval in $\xi$ equal to half the cavity round-trip time $T$. The soliton solution and neutral modes were obtained by discretizing the derivatives and solving the resulting nonlinear equations via Newton's method. }
\label{fig:soliton}
\end{figure}

Solvability of the first-order equations requires the right-hand side of Eq.~(\ref{eq:1storder}) to be orthogonal to the neutral modes of the adjoint operators, leading to the slow-evolution equations
\begin{equation}
\partial_\tau \xi_{1,2}=\pm \frac{u}{2}-K\langle \psi^\dagger_{1,2}\cdot iA^{(0)}_{2,1}\exp(\mp i\delta)\rangle+\langle \psi^\dagger_{1,2}\cdot f_{1,2}\rangle,\label{eq:int}
\end{equation} 
where $\psi^\dagger_{1,2}=\psi^\dagger(\xi-\xi_{1,2})$ denote the adjoint neutral modes satisfying ${\cal L}^\dagger_{1,2}\psi^\dagger_{1,2}=0$ and the normalization condition $\langle\psi^\dagger\cdot\psi\rangle=\int_{0}^{T}(\Re\psi^\dagger\Re\psi+\Im\psi^\dagger\Im\psi)d\xi=1$. These adjoint modes were computed numerically using the approach detailed in Refs.~\cite{vladimirov2018effect,vladimirov2021dissipative}. Real and imaginary parts of the soliton solution $A^{(0)}(\xi)$, neutral mode $\psi(\xi)$, and adjoint neutral mode $\psi^{\dagger}(\xi)$ are shown in Fig.~\ref{fig:soliton}. As can be seen, in addition to the asymmetry, the presence of a TOD causes both the soliton solution and its neutral modes to develop Cherenkov radiation, manifested as a rapidly oscillating tail on one side. This tail can strongly modify the weak interactions between solitons and lead to the emergence of multiple soliton bound states. Because reflection symmetry is broken for $d\neq 0$, the last term in Eq.~(\ref{eq:int}), which accounts for parameter mismatches, is generally nonzero. 

Expanding the inner product in the second term of Eq.~(\ref{eq:int}) gives $K\langle \psi^\dagger_{1,2}\cdot iA^{(0)}_{2,1}\exp(\mp i\delta)\rangle=K\left( C_{1,2}\cos\delta\pm S_{1,2}\sin\delta\right)$ with 
\begin{align}
C_{1,2}=\int_{0}^{T}\left[\Im A^{(0)}(\xi-\xi_{2,1})\Re\psi^\dagger(\xi-\xi_{1,2})-\Re A^{(0)}(\xi-\xi_{2,1})\Im\psi^\dagger(\xi-\xi_{1,2})\right]d\xi\nonumber\\
=\int_{0}^{T}\left[\Im A^{(0)}(\xi)\Re\psi^\dagger(\xi\pm\xi_-)-\Re A^{(0)}(\xi)\Im\psi^\dagger(\xi\pm\xi_-)\right]d\xi
\label{eq:C12}
\end{align}
 and 
\begin{align}
 S_{1,2}=\int_{0}^{T}\left[\Re A^{(0)}(\xi-\xi_{2,1})\Re\psi^\dagger(\xi-\xi_{1,2})+\Im A^{(0)}(\xi-\xi_{2,1})\Im\psi^\dagger(\xi-\xi_{1,2})\right]d\xi\nonumber\\
  =\int_{0}^{T}\left[\Re A^{(0)}(\xi)\Re\psi^\dagger(\xi\pm\xi_-)+\Im A^{(0)}(\xi)\Im\psi^\dagger(\xi\pm\xi_-)\right]d\xi,
\label{eq:S12}
\end{align}
where we have used the $T$-periodicity of the integrand, which allows the integration variable to be shifted arbitrarily. Consequently, the quantities $C_{1,2}$ and $S_{1,2}$ depend only on the soliton separation $\xi_2-\xi_1 = \xi_-$, and not on the individual positions $\xi_{1,2}$ separately. 

Introducing the center-of-mass coordinate $\xi_+=(\xi_1+\xi_2)/2$ and the relative separation $\xi_-=\xi_2-\xi_1$, we can rewrite Eq.~(\ref{eq:int}) as
\begin{eqnarray}
\partial_\tau \xi_{+}=\frac{F_+}{2} +\frac{K}{2}\left[S_+(\xi_-)\sin\delta+C_+(\xi_-)\cos\delta\right],\label{eq:int1_TOD}\\
\partial_\tau \xi_{-}= -u+F_-+K\left[S_-(\xi_-)\sin\delta+C_-(\xi_-)\cos\delta\right],\label{eq:int2_TOD}
\end{eqnarray}
where we have defined $C_{\pm}(\xi_-)=C_2\pm C_1$, $S_{\pm}(\xi_-)=S_2\mp S_1$, and $F_\pm=(F_2\pm F_1)$ with $F_{1,2}=\langle \psi^\dagger_{1,2}\cdot f_{1,2}\rangle$. 

\begin{figure}[thb]
\centering
\includegraphics[width=0.49\linewidth]{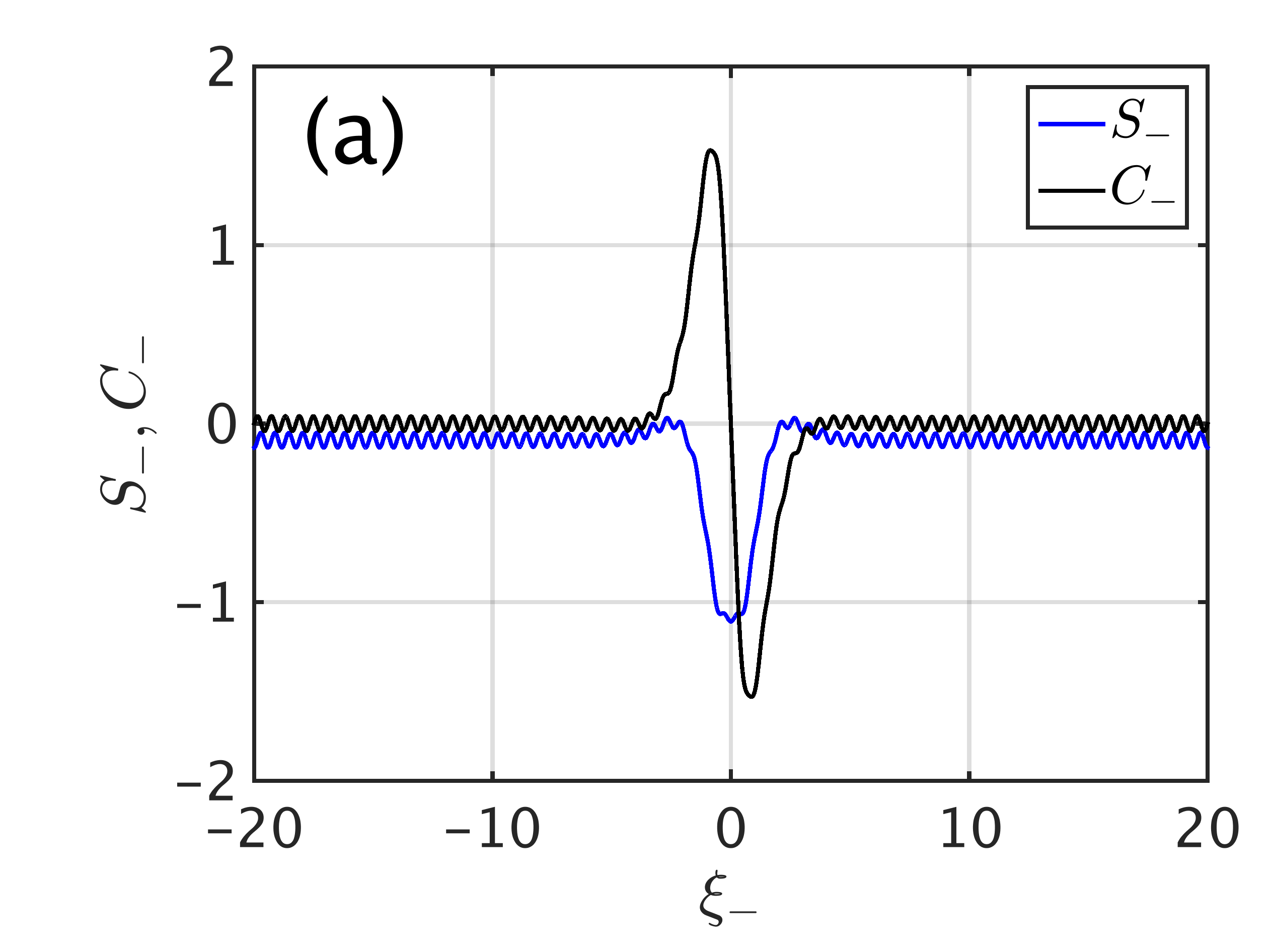}\includegraphics[width=0.49\linewidth]{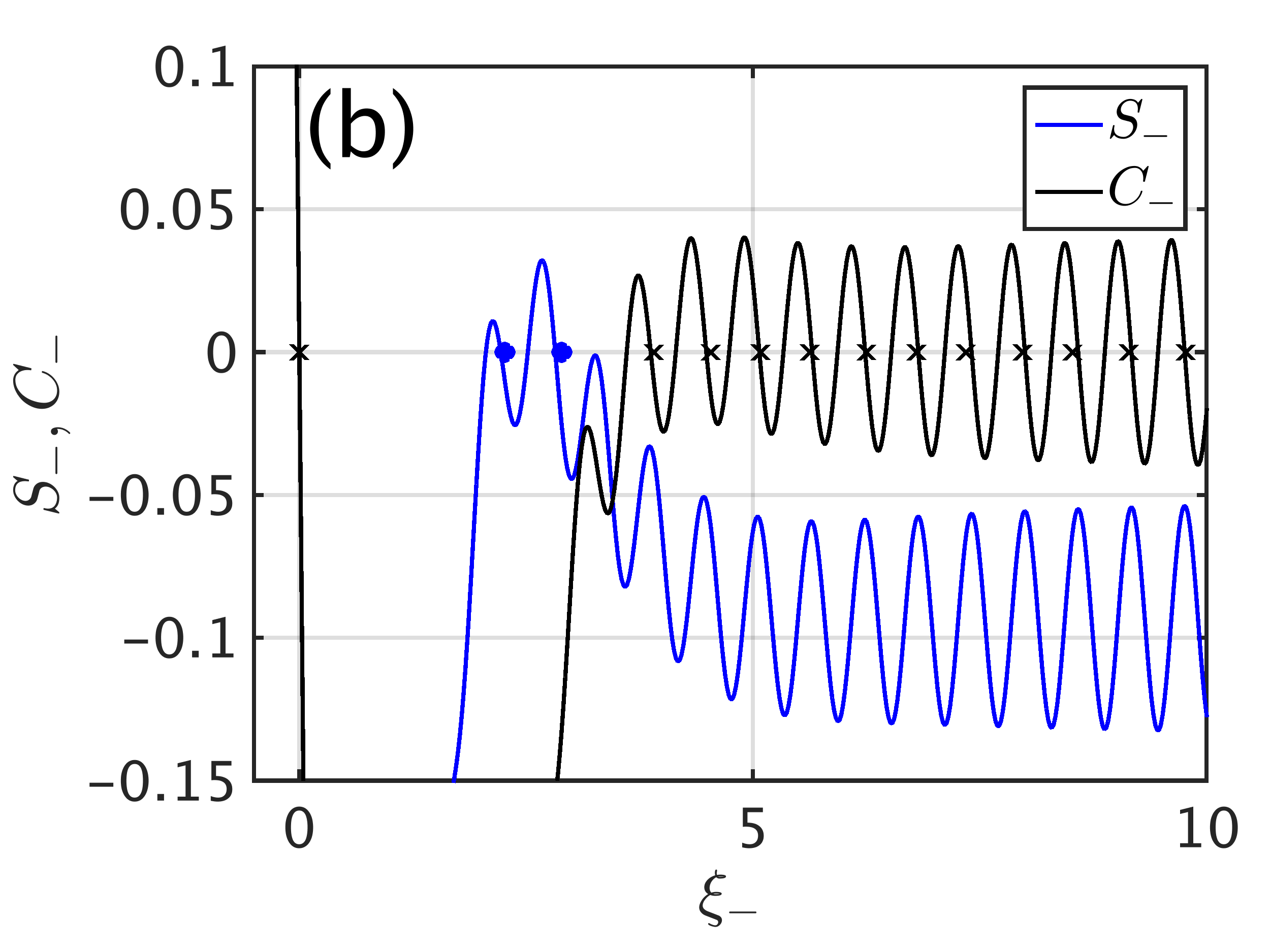}
\includegraphics[width=0.49\linewidth]{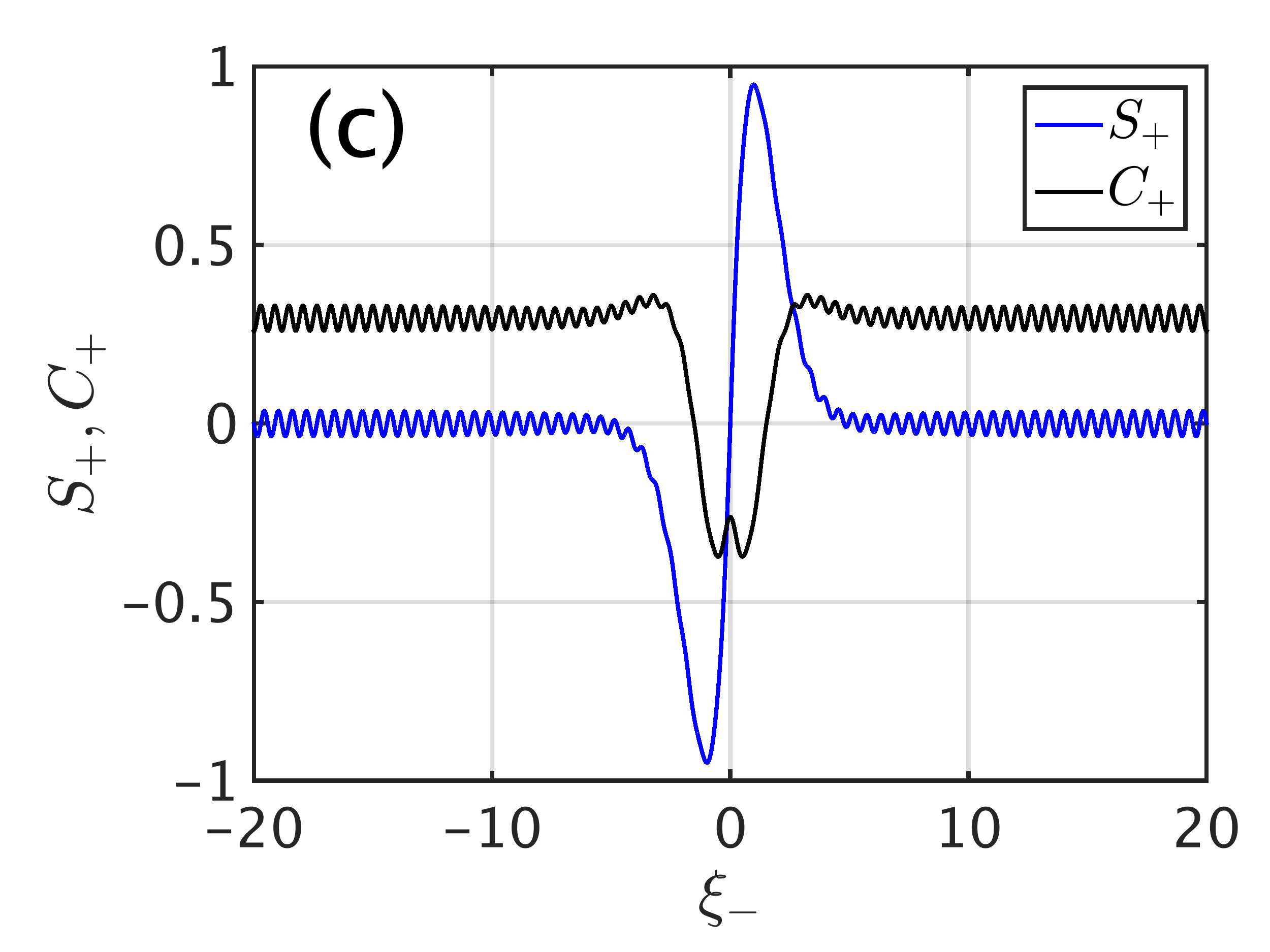}\includegraphics[width=0.49\linewidth]{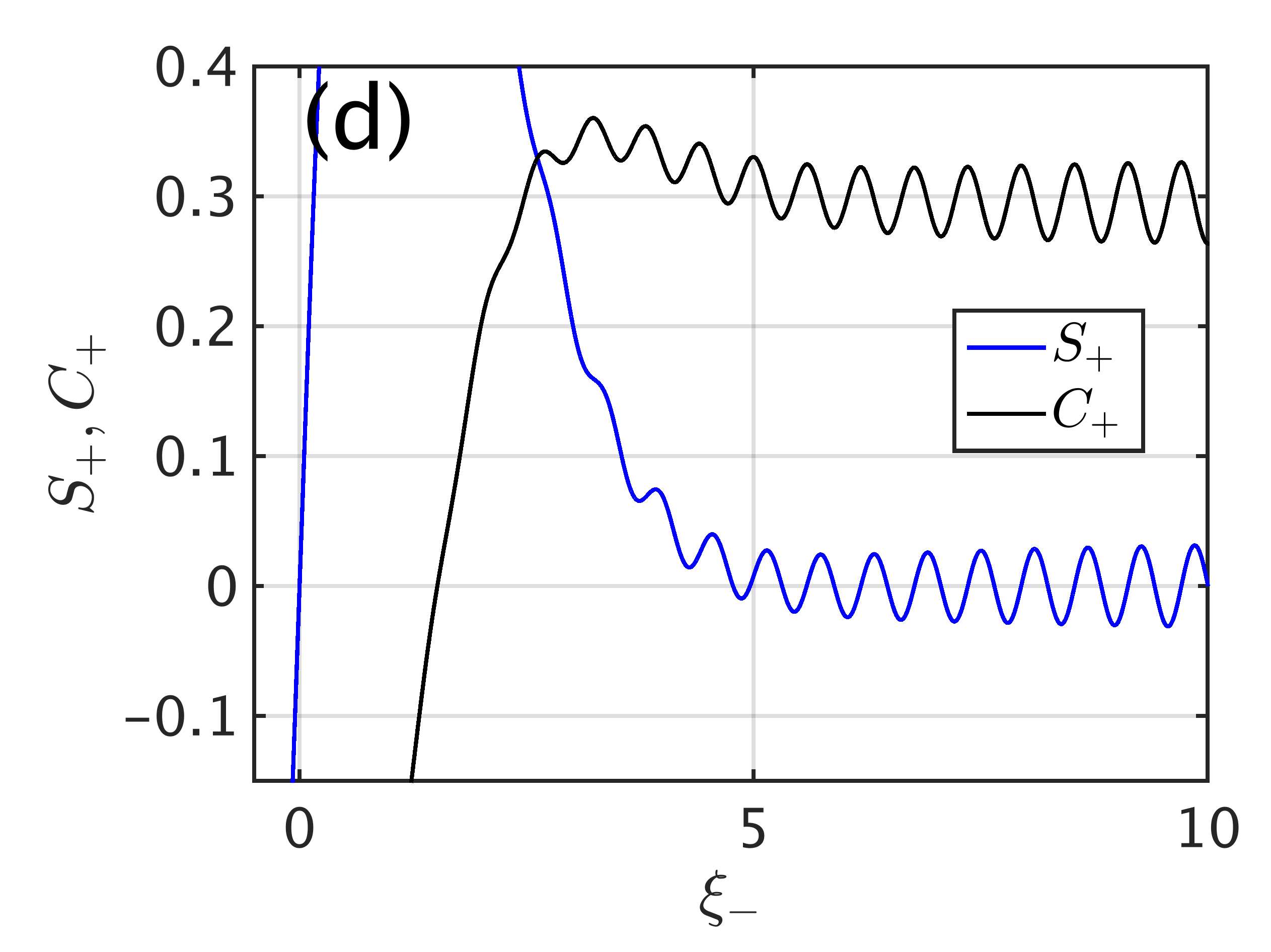}
\caption{Dependence of $S_{\pm}$ (blue) and $C_{\pm}$ (black) on $\xi_-$. The upper panels (a,b) and lower panels (c,d) show $(S_{-}, C_{-})$ and $(S_{+}, C_{+})$, respectively. The right panels (b,d) present a zoom near the coordinate origin to better resolve the zero crossings. Black crosses (blue dots) in panel (b) indicate soliton separations $\xi_-$ in stable soliton clusters calculated by numerical integration of Eqs.~(\ref{eq:LLE_coupled1}) and (\ref{eq:LLE_coupled2}) with $\delta=0$ ($\delta=\pi/2$). The parameters are $\theta = 3.5$, $k=1$, $\eta = 1.9$, $b = 1$, $d=0.1$, and $T = 40.0$.}
\label{fig:SCpm}
\end{figure}

It follows from Eqs.~(\ref{eq:C12}) and (\ref{eq:S12}) that $S_{1,2}$ and $C_{1,2}$ obey the symmetry relations $S_{2,1}(\xi_-)=S_{1,2}(-\xi_-)$ and $C_{2,1}(\xi_-)=C_{1,2}(-\xi_-)$. This implies that $S_{+}$ and $C_{-}$ are odd functions of $\xi_-$, whereas $S_{-}$ and $C_{+}$ are even functions of $\xi_-$. In the absence of TOD, the relations $C_2(\xi_-)=-C_1(\xi_-)$ and $S_2(\xi_-)=-S_1(\xi_-)$ follow from the parity symmetry of the soliton profile and imply that $S_-=C_+=0$. Consequently, for $d=0$, these symmetry relations reduce Eqs.~(\ref{eq:int1_TOD}) and~(\ref{eq:int2_TOD}) to the interaction equations derived in Ref.~\cite{dolinina2026soliton}.  With TOD present, however, these equations acquire corrections proportional to $d$, reflecting the asymmetry of the soliton tails.  When $d\neq 0$, the functions $S_{\pm}$, $C_{\pm}$, and $F_{\pm}$ all depend on the TOD coefficient through the soliton profile and adjoint mode, thereby introducing an additional factor that modifies the interaction. The numerically computed coefficients $S_{\pm}$ and $C_{\pm}$ are shown as functions of $\xi_-$ in Fig.~\ref{fig:SCpm}, plotted in blue and black, respectively.

The separation $\xi_-$ is interpreted as the time lag between the arrivals of the two solitons at the the cross-sections of the microresonators that are closest to each other in the coupling region, see Fig.~\ref{fig:scheme}. For cluster states with fixed $\xi_-$, this lag remains constant from one round trip to the next. In the limiting case $\xi_-=0$, the solitons pass the closest cross-sections simultaneously, resulting in the strongest possible interaction. A cluster characterized by zero separation, $\xi_-=0$, can be referred to as a soliton ``bunching'' state \cite{dolinina2026soliton}. 

Soliton clusters, if they exist, correspond to soliton separations for which $\partial_\tau \xi_{-}=0$; consequently, the right-hand side of Eq.~(\ref{eq:int2_TOD}) must vanish. To identify such clusters and estimate their common drift velocity using Eq.~(\ref{eq:int1_TOD}), it is instructive to examine the dependence of the functions $S_{\pm}$ and $C_{\pm}$ on the soliton separation $\xi_-$. In contrast, the quantities $F_{\pm} \equiv F_2 \pm F_1$ appearing in the interaction equations are independent of $\xi_-$, since the integrals defining $F_1$ and $F_2$ each involve products of the soliton profile and the adjoint neutral mode with matching indices---$1$ for $F_1$ and $2$ for $F_2$. Consequently, $F_-$ introduces an additional relative drift between the solitons, analogous to the effect of the group velocity mismatch parameter $u$, but fixed for a given soliton shape. Similarly, $F_+$ in Eq.~(\ref{eq:int1_TOD}) gives rise to a common drift of both solitons, independent of their separation. 
In view of these observations, we restrict our analysis to the case $F_1 = F_2 = 0$, meaning that the two cavities are identical except for the small cavity group velocity mismatch $u$.

\begin{figure}[htbp]
  \centering
  \begin{minipage}{0.50\textwidth}
    \centering
   \includegraphics[width=\linewidth]{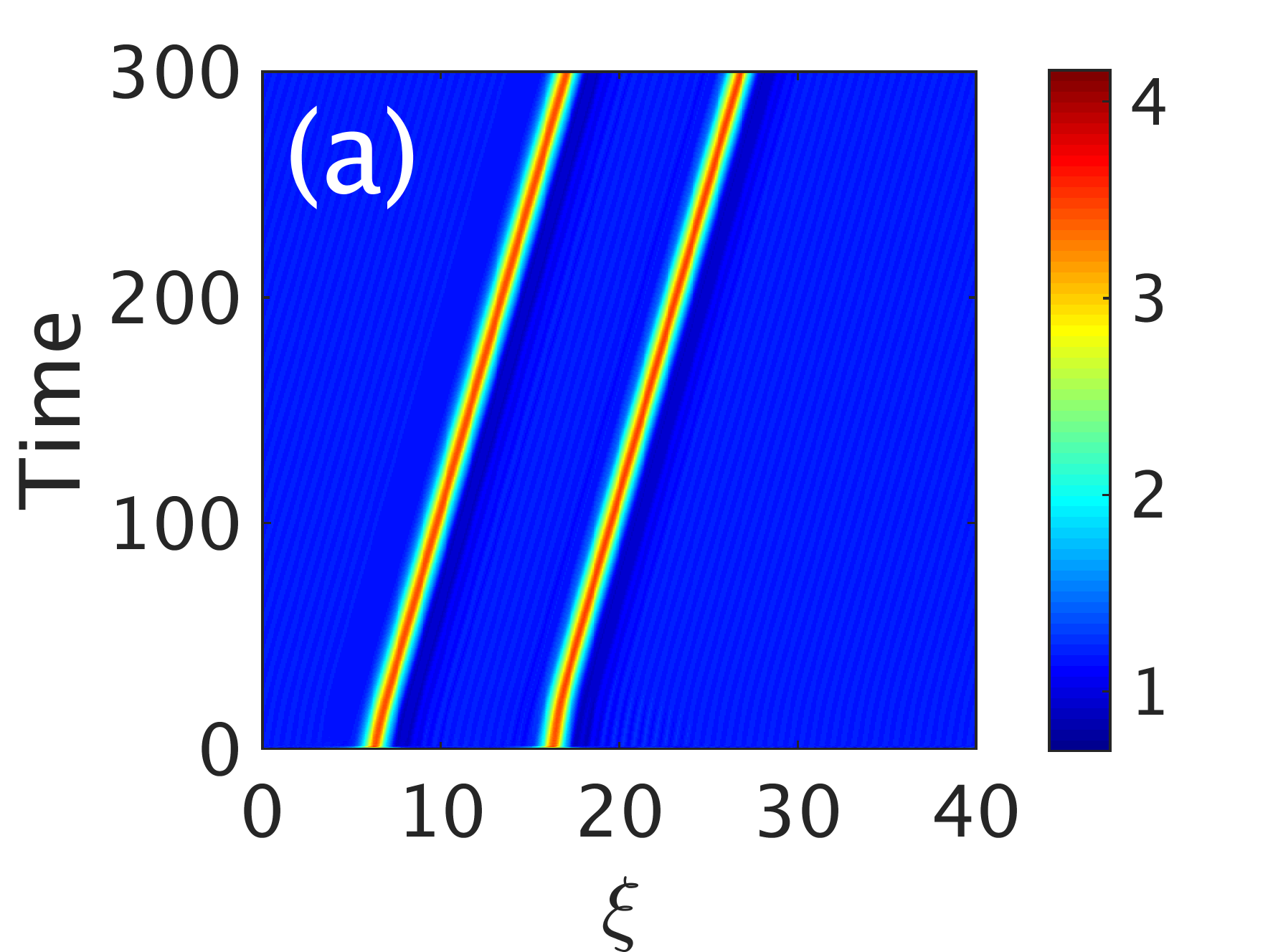}
  \end{minipage}
  \hfill
  \begin{minipage}{0.48\textwidth}
    \centering
    \includegraphics[width=\linewidth]{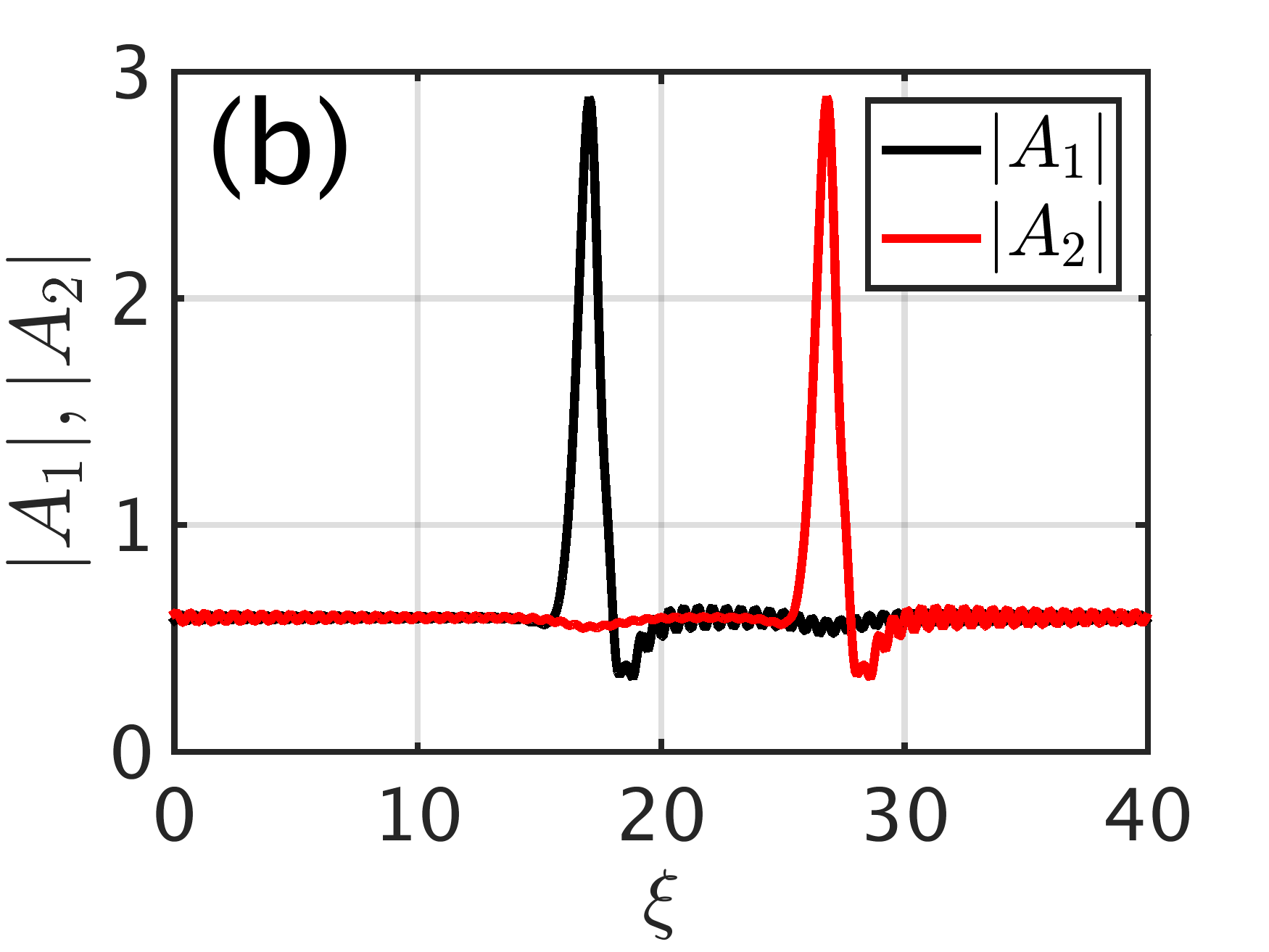}
  \end{minipage}
  \vspace{0.2cm}

  \begin{minipage}{0.48\textwidth}
    \centering
  \end{minipage}
  \hfill
  \begin{minipage}{0.48\textwidth}
    \centering
  \end{minipage}
\caption{Soliton cluster calculated using the pseudospectral integration method applied to Eqs.~(\ref{eq:LLE_coupled1}) and (\ref{eq:LLE_coupled2}) for zero injection phase difference, $\delta=0$. (a) Time evolution of the total intensity $|A_1|+|A_2|$, starting from the initial two Gaussian distributions centered at $\xi_{1}=16.0$ and $\xi_{2}=24.0$. For better visualization of the interaction, the soliton drift is partially compensated for in the present and all analogous figures below. (b) Field amplitudes $|A_1|$ and $|A_2|$ in the cluster state. Parameters: $V=0$, $\eta_1=\eta_2=1.9$, $\theta_1=\theta_2=3.5$, $k_1=k_2=1$, $b_1=b_2=1$, $d_1=d_2=0.1$, $\kappa=0.1$, and $T=40.0$.}
  \label{fig:delta0}
\end{figure}

We begin by examining Eqs.~(\ref{eq:int1_TOD}) and (\ref{eq:int2_TOD}) for the case of identical cavities ($F_1=F_2=0$) with no optical length mismatch ($u=0$). We consider several limiting cases corresponding to injection-field phase differences $\delta=0,\pm\pi,\pm\pi/2,\pm\pi/4$.  For $\delta=0,\pm\pi$, we have $\sin{\delta}=0$, and the soliton clusters are determined by the condition $C_-=0$. As seen from Figs.~\ref{fig:SCpm}(a,b), this case features a ``bunching'' cluster with zero separation between the solitons ($\xi_-=0$), along with multiple clusters whose stability alternates depending on the sign of the slope of the $C_-$-curve. The existence of multiple clusters with alternating stability is typical of two-soliton interactions in the LLE with higher-order dispersion (see, e.g., \cite{parra2017interaction,vladimirov2018effect,vladimirov2021dissipative}). As follows from Eq.~(\ref{eq:int2_TOD}), replacing $\delta=0$ with $\delta=\pm\pi$ interchanges the stability of all clusters, converting stable ones into unstable and vice versa. The positions of stable clusters calculated for $\delta=0$ by numerical integration of the coupled LLEs (\ref{eq:LLE_coupled1}) and (\ref{eq:LLE_coupled2}) using a pseudo-spectral method are shown by black crosses in Fig.~\ref{fig:SCpm}(b), and show reasonable agreement with those predicted by the condition $C_-=0$. One of these stable clusters is illustrated in Fig.~\ref{fig:delta0}.

\begin{figure}[htbp]
  \centering
  \begin{minipage}{0.50\textwidth}
    \centering
   \includegraphics[width=\linewidth]{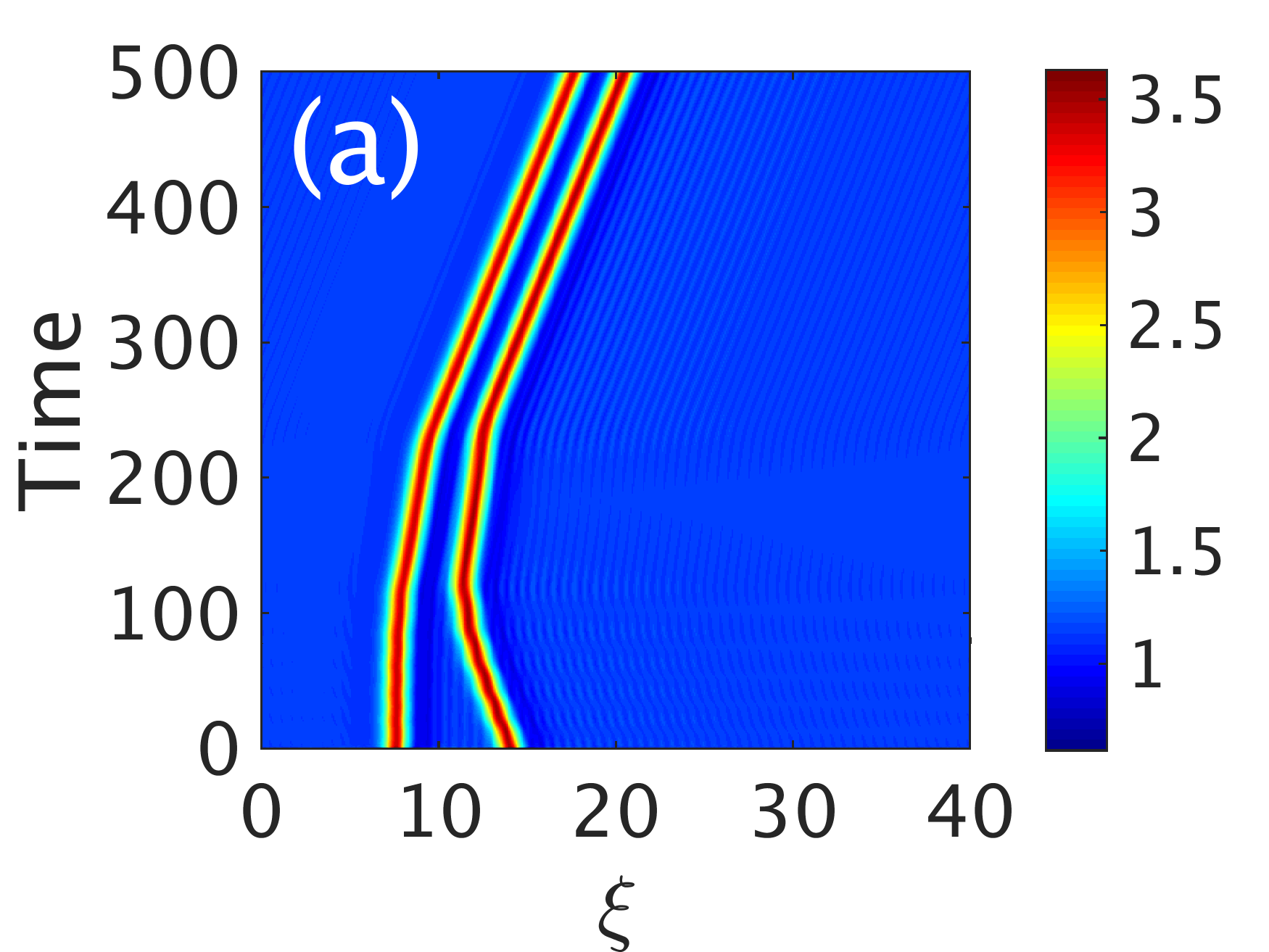}
  \end{minipage}
  \hfill
  \begin{minipage}{0.48\textwidth}
    \centering
    \includegraphics[width=\linewidth]{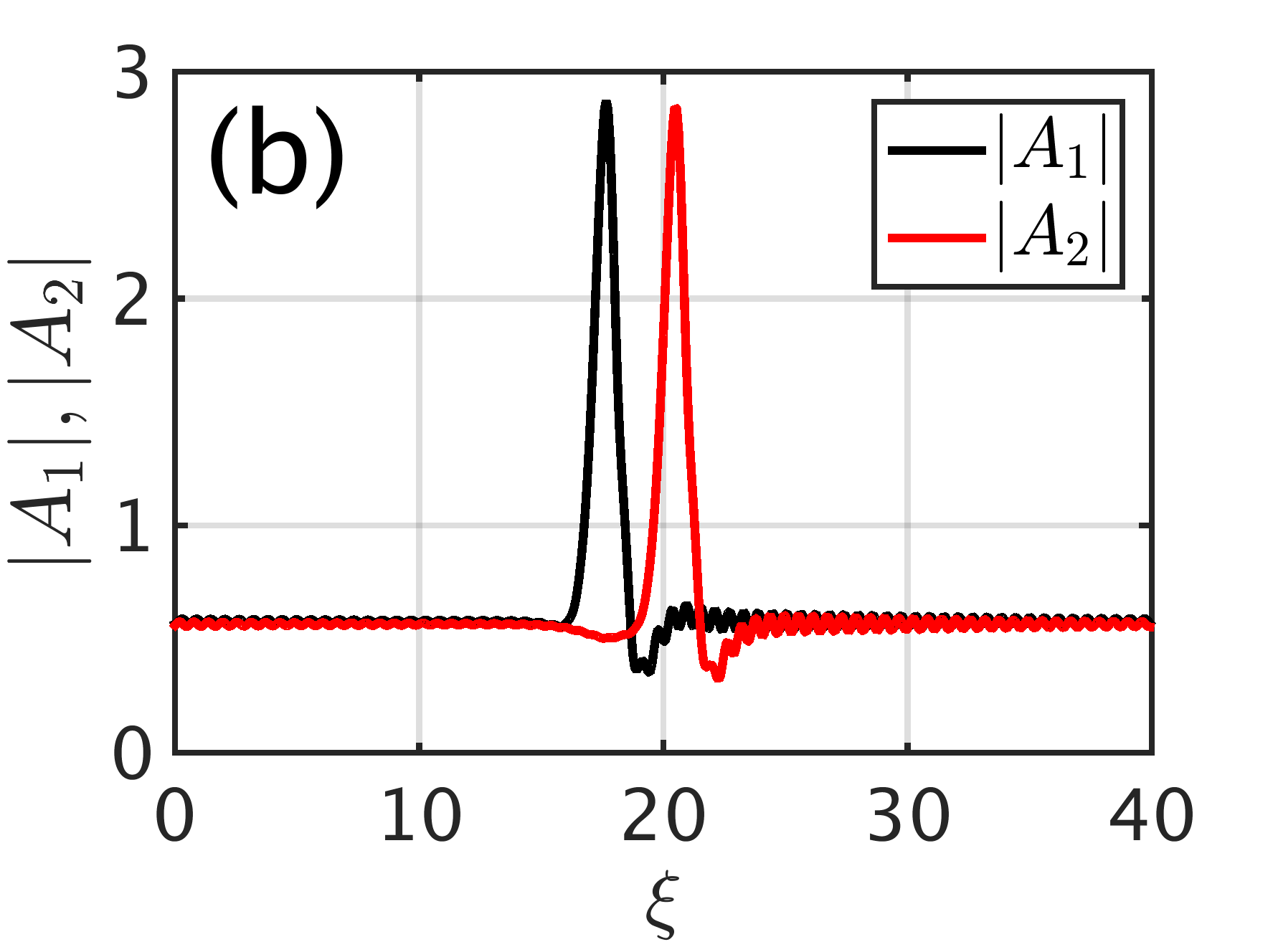}
  \end{minipage}
\caption{Soliton cluster calculated by split-step integration of Eqs.~(\ref{eq:LLE_coupled1}) and (\ref{eq:LLE_coupled2}) with injection phase difference $\delta=\pi/2$. This cluster corresponds to the right blue point in Fig~\ref{fig:SCm}(b). (a) Time evolution of the total intensity $|A_1|+|A_2|$ starting from the initial two Gaussian distributions centered at $\xi_{1}=5.0$ and $\xi_{2}=15.0$. (b) Field amplitudes $|A_1|$ and $|A_2|$ in the cluster state. Other parameters are as in Fig.~\ref{fig:delta0}.}
  \label{fig:delta05}
\end{figure}

Next, consider the case $\delta=\pi/2$, for which the existence condition for clusters with a fixed soliton separation $\xi_-$ reduces to $S_-=0$. In this case, it follows from Figs.~\ref{fig:SCpm}(a,b) that only two stable soliton clusters with positive separations around $\xi\approx 3$ are possible in two identical cavities at zero group velocity mismatch $V=u=0$. Blue dots in Fig.~\ref{fig:SCpm}(b) indicate the positions of these clusters, calculated by numerical integration of Eqs.~(\ref{eq:LLE_coupled1}) and (\ref{eq:LLE_coupled2}). The stable cluster with the maximal soliton separation is illustrated in Fig.~\ref{fig:delta05}. In contrast to Fig.~\ref{fig:delta0}, where the initial soliton separation is close to that of one of the cluster states, the soliton separation here decreases substantially before reaching its stationary value. As seen in Fig.~\ref{fig:delta05}(a), before settling into the cluster state corresponding to the rightmost blue dot in Fig.~\ref{fig:SCpm}(b), the solution spends some time near the local maximum of the blue $S_-$ line located to the right of this dot.


It follows from Fig.~\ref{fig:SCpm}(b) and Eq.~(\ref{eq:int2_TOD}) that, although only two stable clusters with $\xi_2>\xi_1$ are possible for $u=0$ and $\delta=\pi/2$, the number of stable clusters can be increased by an appropriate choice of the group-velocity mismatch, $u\neq0$. Indeed, for nonzero $u$ and $F_-=0$, clusters with stationary $\xi_-$ are determined by the condition
$S_-\sin\delta+C_-\cos\delta-u/K=0$,
rather than by
$S_-\sin\delta+C_-\cos\delta=0$.
For $\delta=\pi/2$, this condition reduces to $S_--u/K=0$, corresponding to a vertical shift of the $S_-$ curve by $u/K$ along the $y$-axis. As it is seen from Fig.~\ref{fig:SCpm}(b), an appropriate shift, such as that corresponding to $u/K=-0.1$, results in multiple intersections of $S_--u/K$ with the $\xi_-$-axis. Consequently, multiple stable soliton clusters can exist at $\delta=\pi/2$ for a suitably chosen cavity group-velocity mismatch. As the magnitude of the group-velocity mismatch is increased further, it follows from Fig.~\ref{fig:SCpm}(a) that the clusters disappear through a sequence of saddle-node bifurcations, eventually leaving a single stable cluster, which itself vanishes at sufficiently large negative $u$. A similar scenario involving saddle-node bifurcations arises for $\delta=0$, with the key difference that multiple stable soliton clusters can exist even in the absence of group-velocity mismatch ($u=0$). For $\delta=-\pi/2$, the condition for stationary soliton separations becomes $-S_--u/K=0$. Hence, in this case, multiple soliton clusters arise for positive $u/K$, for example, at $u/K=0.1$. 

\begin{figure}[thb]
\centering
\includegraphics[width=0.49\linewidth]{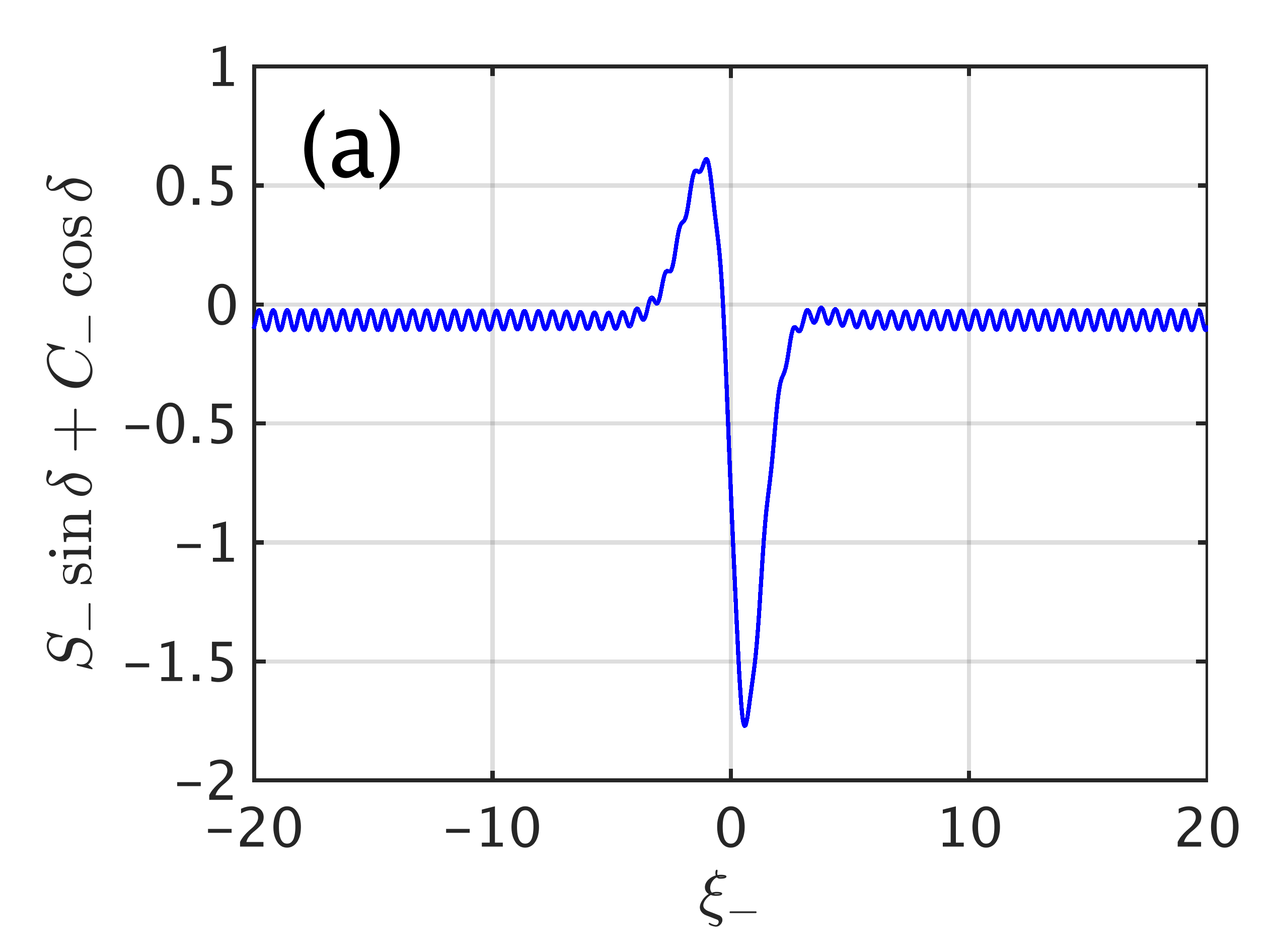}\includegraphics[width=0.49\linewidth]{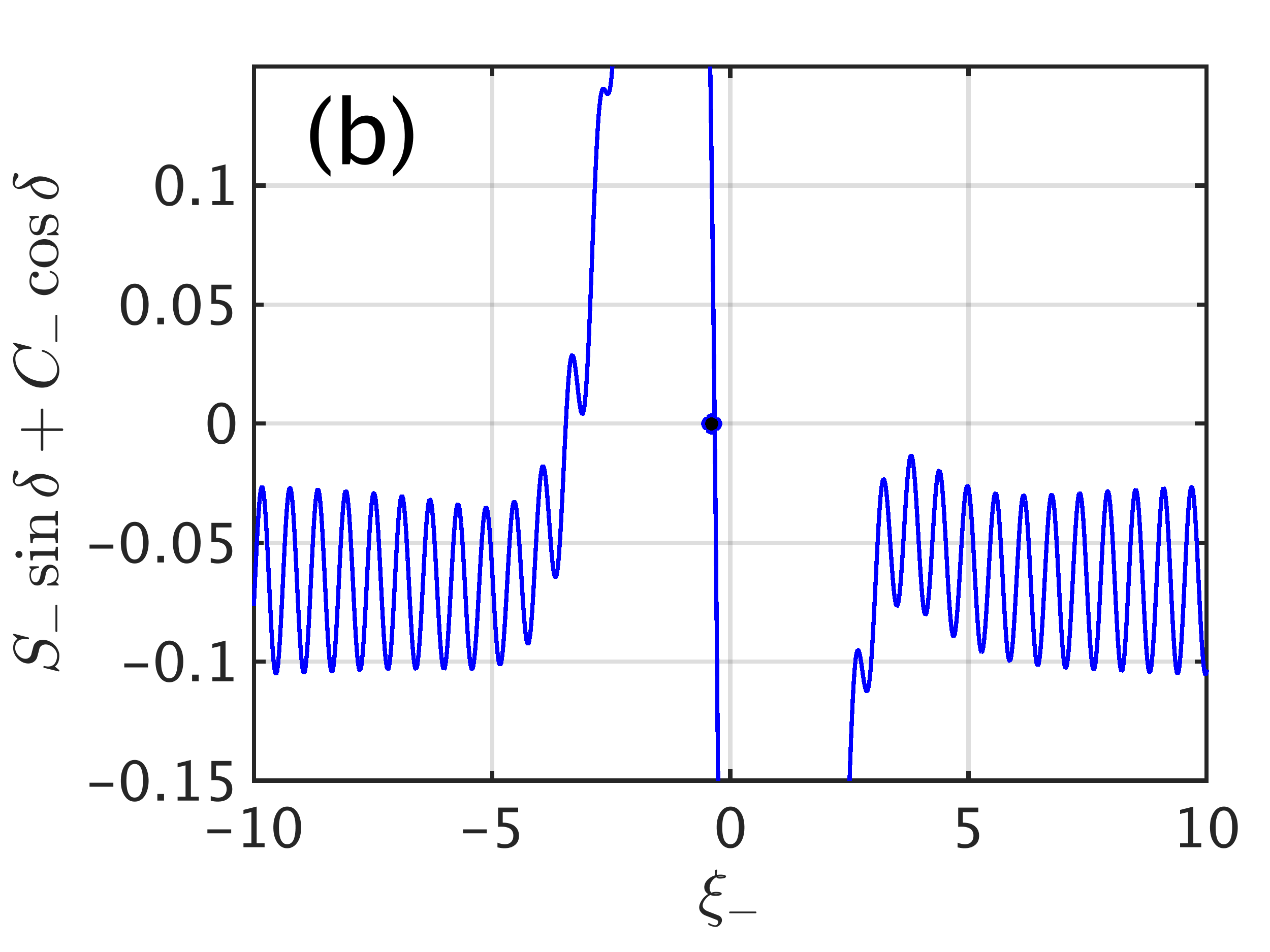}
\caption{Dependence of $S_{\pm}\sin{\delta}+C_{\pm}\cos{\delta}$ with $\delta=\pi/4$ on soliton separation $\xi_-$. The right panel (b) presents a zoom near the coordinate origin to better resolve the zero crossings. Blue dot in panel (b) indicate soliton separation $\xi_-$ in stable soliton cluster calculated by numerical integration of Eqs.~(\ref{eq:LLE_coupled1}) and (\ref{eq:LLE_coupled2}) with $\delta=\pi/4$. Other parameters are as in Fig.~\ref{fig:SCpm}.}
\label{fig:SCm}
\end{figure}

\begin{figure}[htbp]
  \centering
  \begin{minipage}{0.48\textwidth}
    \centering
   \includegraphics[width=\linewidth]{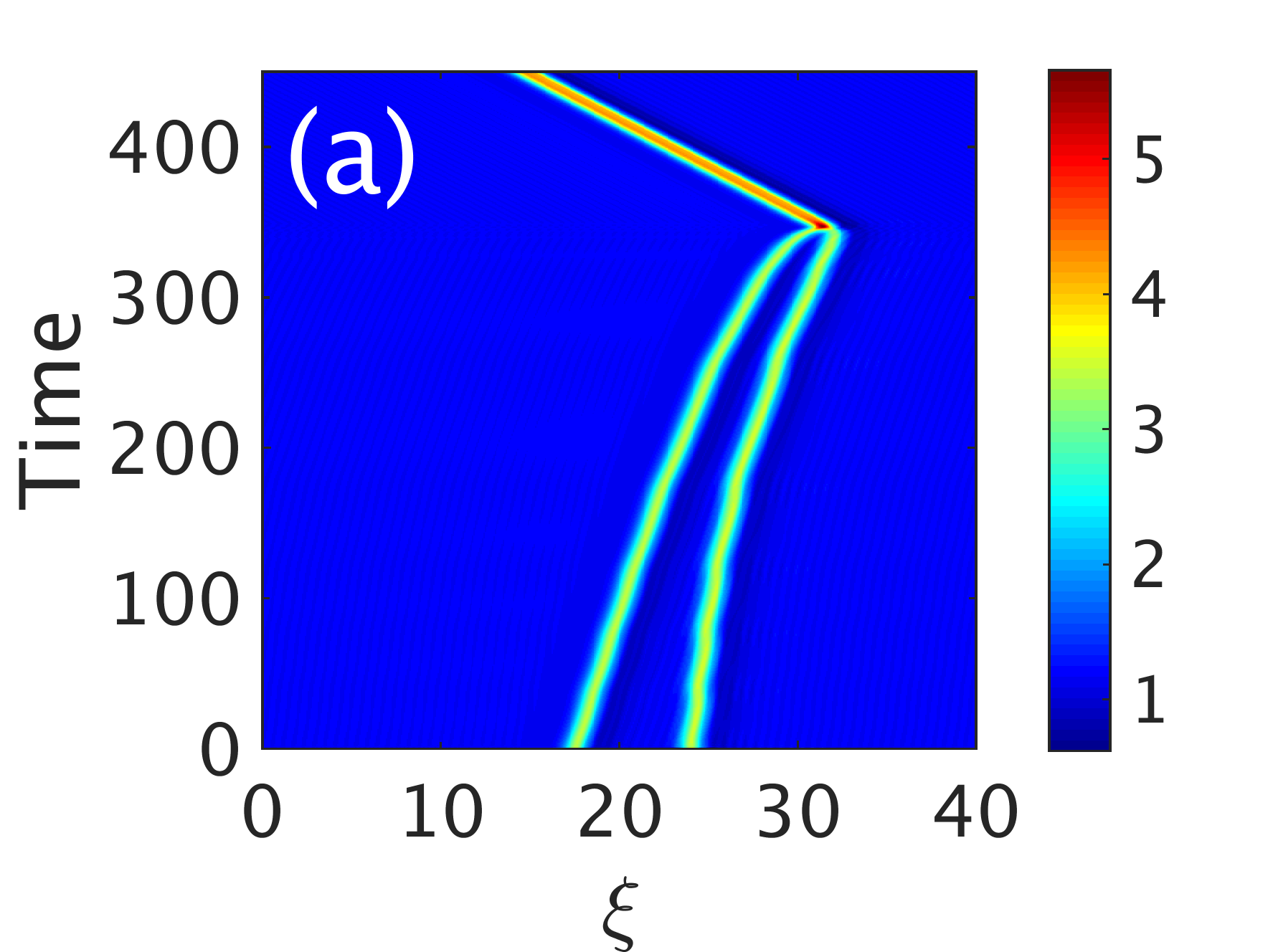}
  \end{minipage}
  \hfill
  \begin{minipage}{0.48\textwidth}
    \centering
    \includegraphics[width=\linewidth]{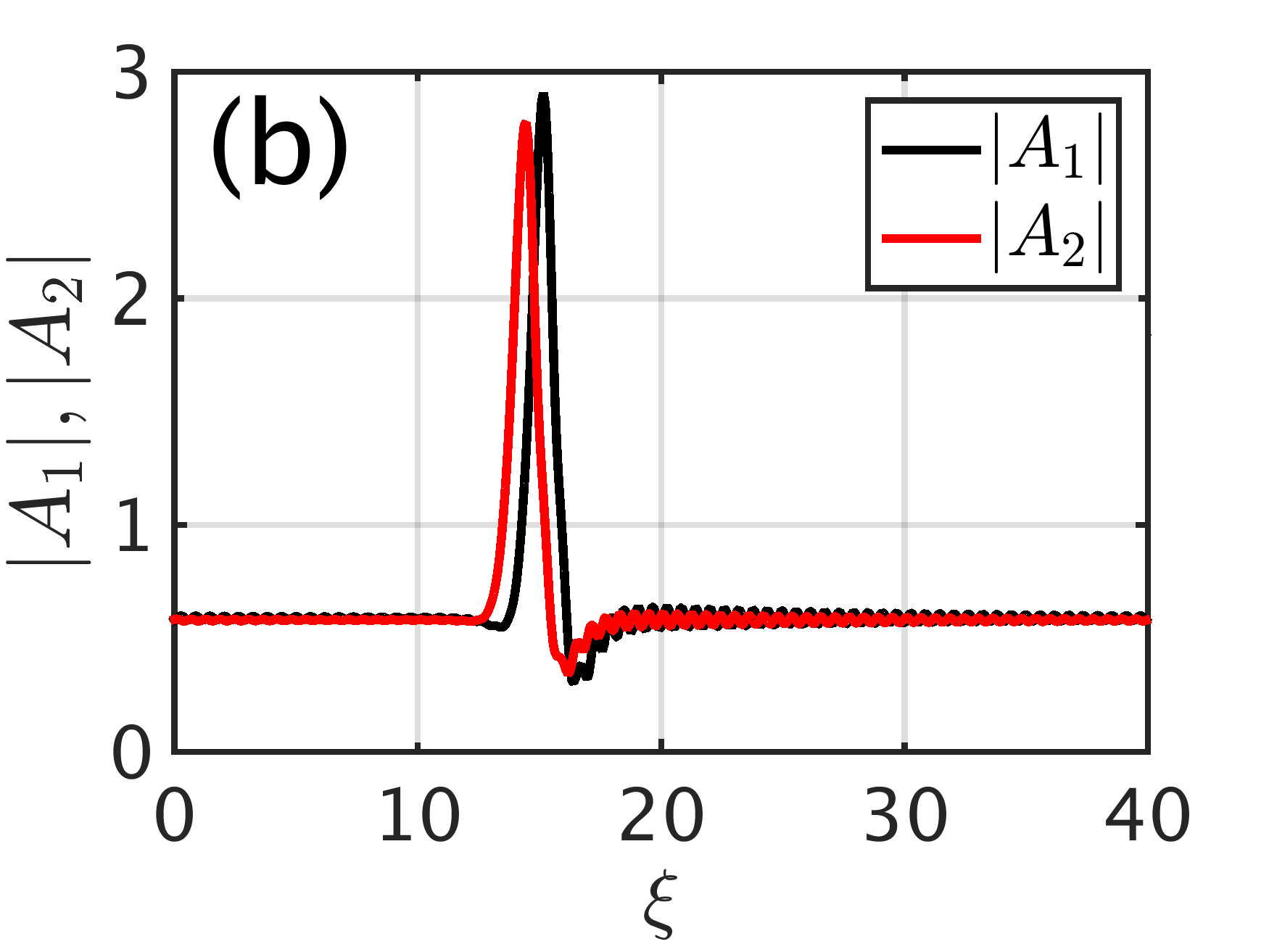}
  \end{minipage}
  
\caption{Field amplitudes of a stable soliton cluster calculated by split-step integration of Eqs.~(\ref{eq:LLE_coupled1}) and (\ref{eq:LLE_coupled2}) with injection phase difference $\delta=\pi/4$. (a) Time evolution of the total intensity $|A_1|+|A_2|$ starting from the initial two Gaussian distributions centered at $\xi_{1}=5.0$ and $\xi_{2}=15.0$. (b) Field amplitudes $|A_1|$ and $|A_2|$ in the cluster state. $\delta=\pi/4$ Other parameters are as in Fig.~\ref{fig:delta0}.}
  \label{fig:delta025}
\end{figure}

Finally, let us consider the intermediate case $\delta=\pi/4$, for which $\sin\delta=\cos\delta=1/\sqrt{2}$. For $u=0$, stationary clusters are determined by the condition
$S_-\sin\delta+C_-\cos\delta=0$.
At $\delta=\pi/4$, the quantity $S_-\sin\delta+C_-\cos\delta$ is a linear combination of an odd and an even function of $\xi_-$ and is therefore neither odd nor even. Its zeros correspond to the intersections of the blue curve in Fig.~\ref{fig:SCm} with the $\xi_-$-axis. As shown in the figure, for $\delta=\pi/4$, only one stable cluster exists, corresponding to an intersection with negative slope at a small negative soliton separation, $\xi_-<0$. The formation of this stable cluster and the corresponding soliton profiles are illustrated in Fig.~\ref{fig:delta025}. In Fig.~\ref{fig:delta025}(b), the second soliton is positioned to the left of the first within the cluster, corresponding to a negative soliton separation, $\xi_-=\xi_2-\xi_1<0$. Figure~\ref{fig:delta025}(a) shows that the soliton separation initially takes a positive value, crosses zero, and subsequently settles at a small negative value. Thus, this cluster can be interpreted as evolving from the ``bunching'' state with zero soliton separation \cite{dolinina2026soliton}, which exists for $\delta=0$ and is indicated by the leftmost black cross in Fig.~\ref{fig:SCpm}(b). Similar to the case $\delta=\pi/2$, multiple stable soliton clusters can also be obtained for $\delta=\pi/4$ by introducing an appropriate nonzero group-velocity mismatch $u$, for example, $u=-0.05$. Further details on the role of group-velocity mismatch in the formation of soliton clusters are provided in the next section.

\section{Effect of group velocity mismatch: Numerical results}

\begin{figure}[htbp]
  \centering
  \begin{minipage}{0.48\textwidth}
    \centering
   \includegraphics[width=\linewidth]{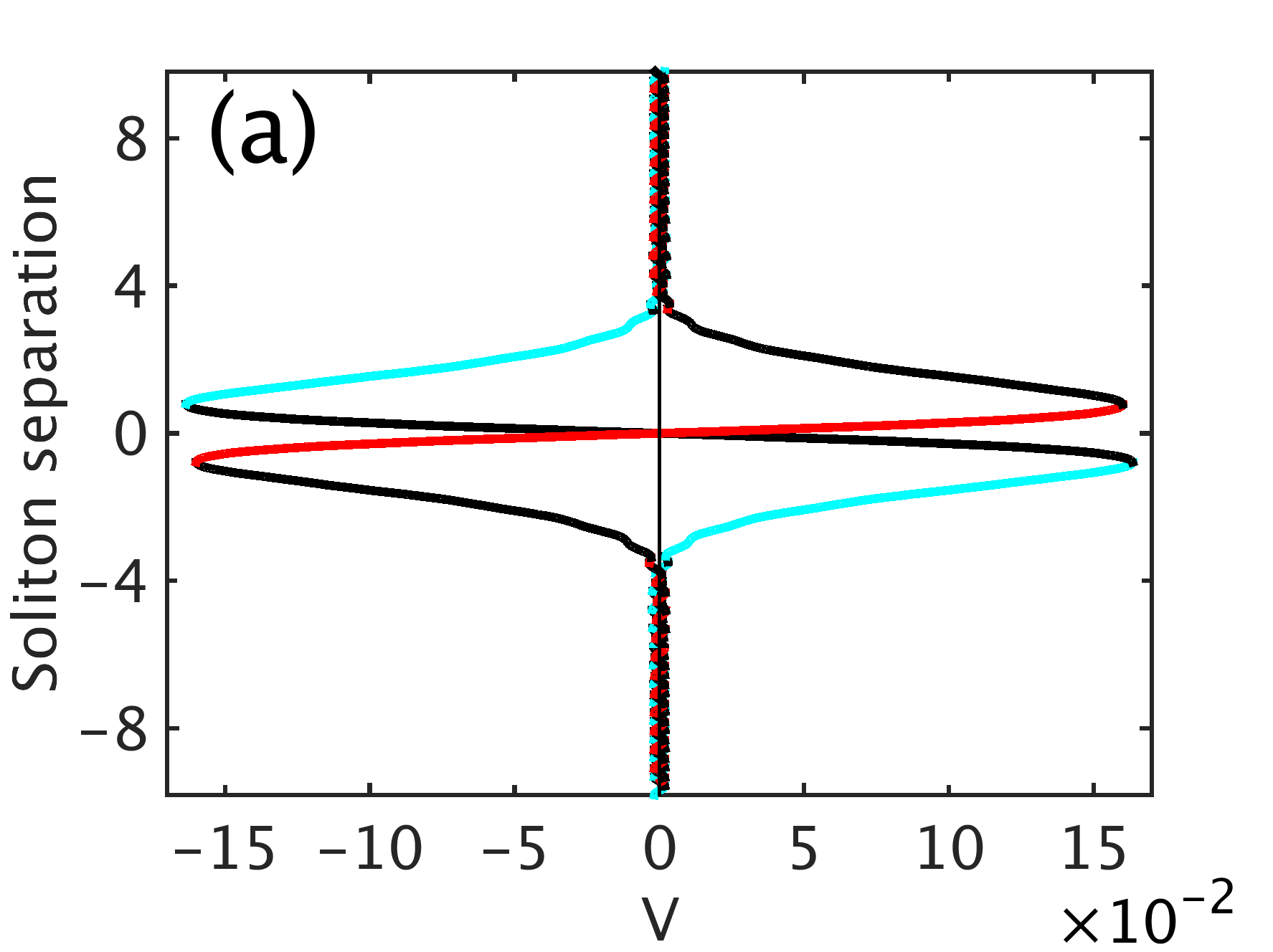}
  \end{minipage}
  \hfill
  \begin{minipage}{0.48\textwidth}
    \centering
    \includegraphics[width=\linewidth]{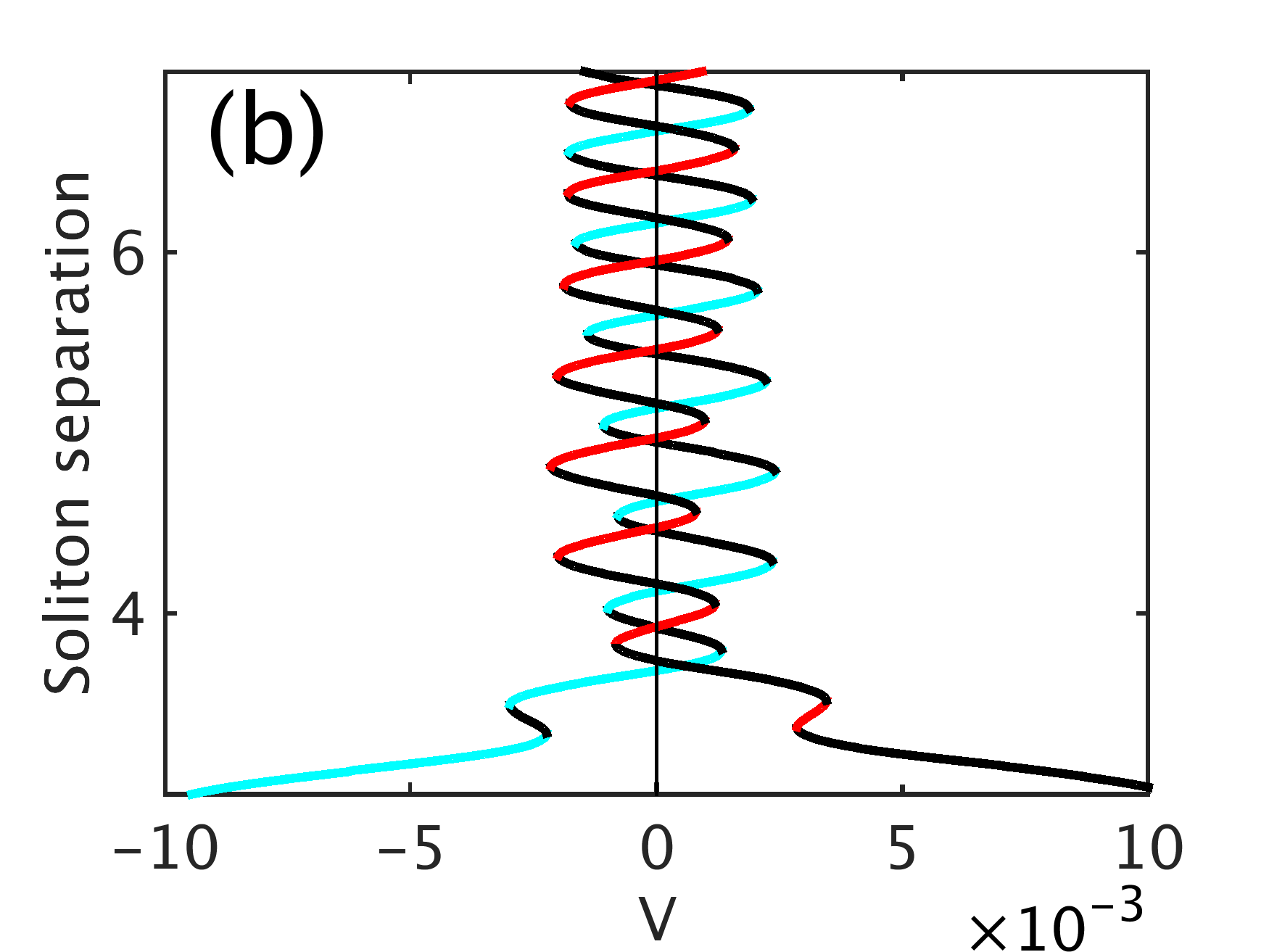}
  \end{minipage}
\caption{Soliton separation $\xi_-$ as a function of the group-velocity mismatch $V$ for the cluster branch, obtained from Eqs.~(\ref{eq:LLE_coupled1}) and (\ref{eq:LLE_coupled2}) using a path-following algorithm (a). Black and cyan (red) indicate the stable and unstable portions of the soliton branch for $\delta=0$ ($\delta=\pi$), respectively. Panel (b): magnified view of the vicinity of the $V=0$ axis. The vertical line indicates $V=0$. Here, $\kappa=0.1$; all other parameters are as in Fig.~\ref{fig:delta0}.}
  \label{fig:distance0}
\end{figure}

In this section, we present the results of a parameter continuation of soliton-cluster solutions obtained numerically using a path-following algorithm. We investigate the dependence of the soliton separation $\xi_-$ within a cluster on the cavity group-velocity mismatch parameter $V$. The two-soliton-cluster branches corresponding to in-phase ($\delta=0$) and antiphase ($\delta=\pi$) pumping beams are shown in Fig.~\ref{fig:distance0}. Stable solutions are indicated by black lines, while unstable cluster solutions for $\delta=0$ and $\delta=\pi$ are shown in cyan and red, respectively. 

For $F_-=\sin\delta=0$, the symmetry property $C_-(-\xi)=-C_-(\xi)$ implies that the asymptotic equation (\ref{eq:int2_TOD}) with $F_-=0$, derived in the limit $\kappa\to0$, is invariant under the transformation $(u,\xi_-)\to(-u,-\xi_-)$. This symmetry reflects the exact symmetry $(V,A_1,A_2)\to(-V, A_2,A_1)$ of the coupled LLEs (\ref{eq:LLE_coupled1}) and (\ref{eq:LLE_coupled2}) for identical coefficients and $\delta=0$ or $\pi$. Consequently, each of the two branches shown in Fig.~\ref{fig:distance0} is symmetric under the transformation $(V,\xi_-)\to(-V,-\xi_-)$. Furthermore, Eq.~(\ref{eq:int2_TOD}) with $F_-=0$ is invariant under the transformation $(u,\tau)\to(-u,-\tau)$, provided that $\delta=0$ and $\delta=\pi$ are interchanged simultaneously.  However, this latter symmetry has no counterpart in the original coupled LLEs (\ref{eq:LLE_coupled1}) and (\ref{eq:LLE_coupled2}). This explains the slight asymmetry between the two branches in Fig.~\ref{fig:distance0}, corresponding to $\delta=0$ and $\delta=\pi$, which are obtained at a finite coupling strength of $\kappa=0.1$.

As seen in Fig.~\ref{fig:distance0}, and in agreement with our asymptotic theory, the two branches exhibit multiple stable and unstable soliton-cluster states for sufficiently small absolute values of the group-velocity mismatch $\abs{V}$. In addition, stable and unstable ``bunching'' states with $\xi_-=0$ are observed for $\delta=0$ and $\delta=\pi$, respectively. As the absolute value of $V$ increases, the soliton-cluster branches disappear through successive saddle-node bifurcations, leaving only a single stable cluster. At larger $\abs{V}$, this remaining cluster undergoes a final saddle-node bifurcation through collision with an unstable cluster, resulting in the disappearance of both clusters.

\begin{figure}[htbp]
    \centering
    \begin{minipage}[t]{0.48\textwidth}
        \centering
        \includegraphics[width=\linewidth]{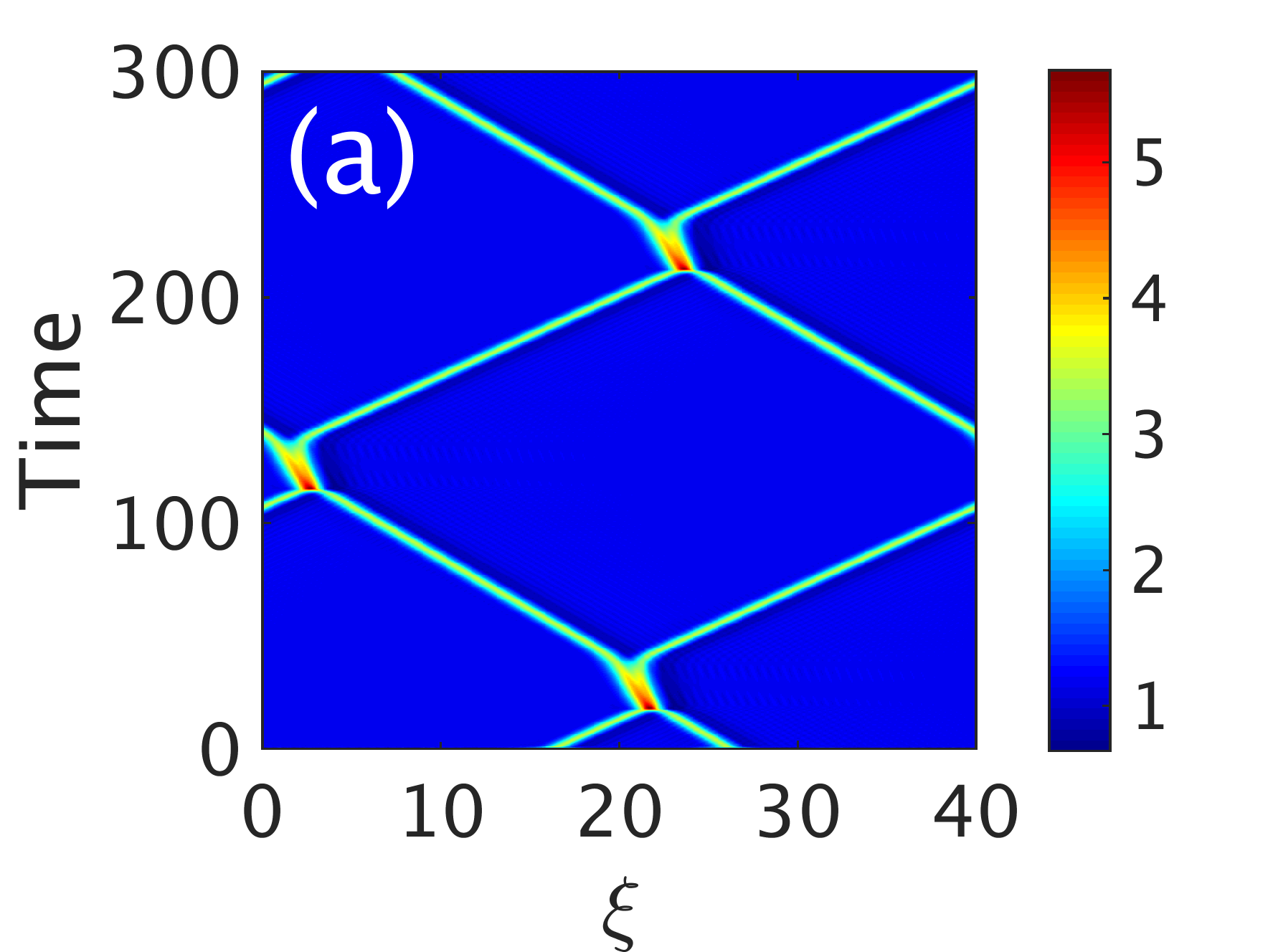}
    \end{minipage}
    \hfill
    \begin{minipage}[t]{0.48\textwidth}
        \centering
        \includegraphics[width=\linewidth]{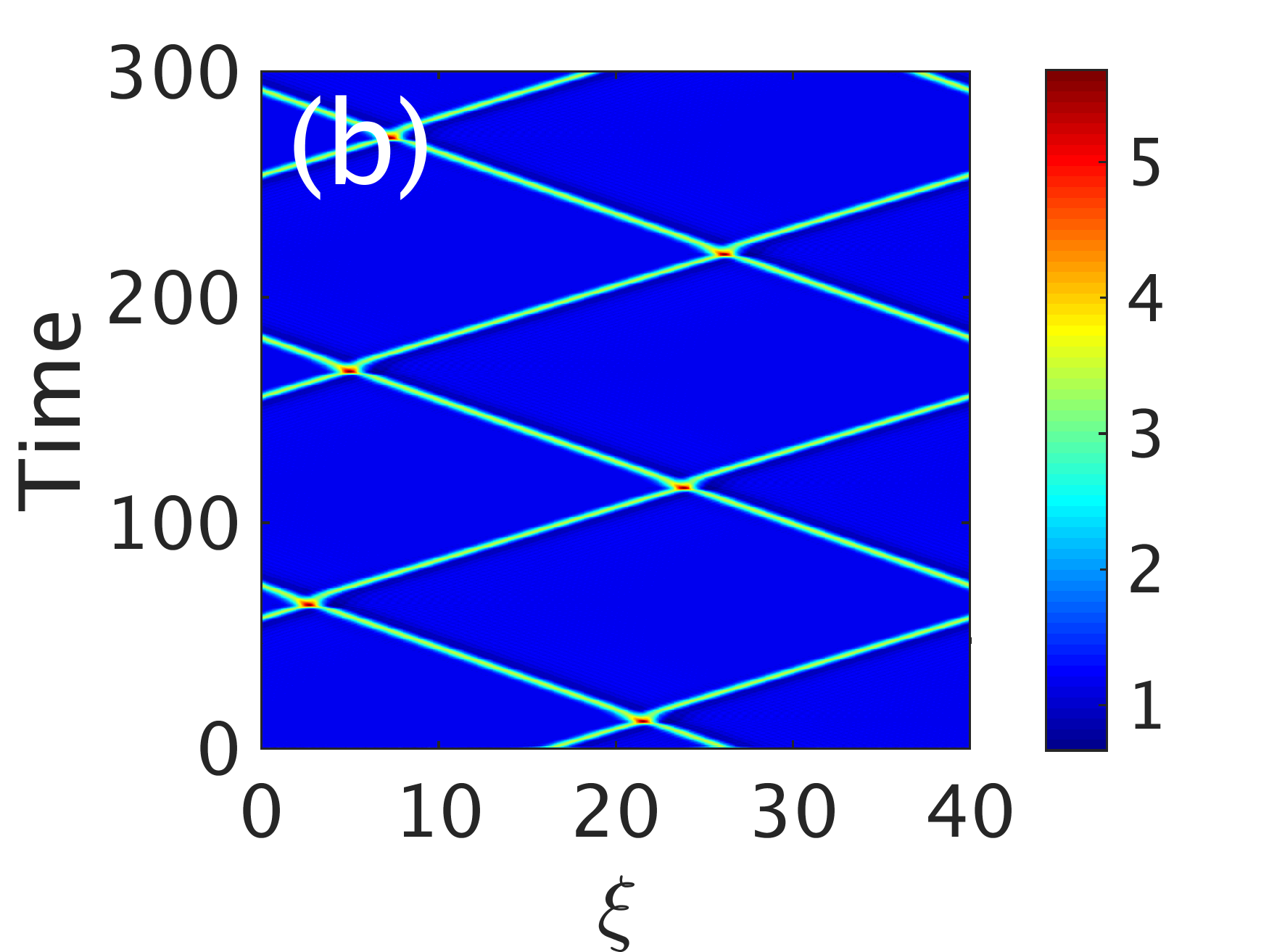}
    \end{minipage}
\caption{Soliton interaction above the uppermost saddle-node bifurcation point, obtained by numerical integration of Eqs.~(\ref{eq:LLE_coupled1}) and (\ref{eq:LLE_coupled2}) with $\delta=0$. Results are shown for $V=0.17$ in (a) and $V=0.27$ in (b); all other parameters are as in Fig.~\ref{fig:distance0}.}
  \label{fig:map0}
\end{figure}

The dynamics of the interacting solitons beyond the final saddle-node bifurcation is shown in Fig.~\ref{fig:map0}. The solitons propagate in opposite directions and undergo periodic collisions, accompanied by shifts in their positions [Fig.~\ref{fig:map0}(a)]. These shifts result from the temporary slowing of the solitons during their interaction near the location of the former saddle-node bifurcation. As the distance from the saddle-node bifurcation increases, the relative velocity increases, while the collision-induced position shift decreases, as shown in Fig.~\ref{fig:map0}(b).

\begin{figure}[htbp]
  \centering
  \begin{minipage}{0.48\textwidth}
    \centering
   \includegraphics[width=\linewidth]{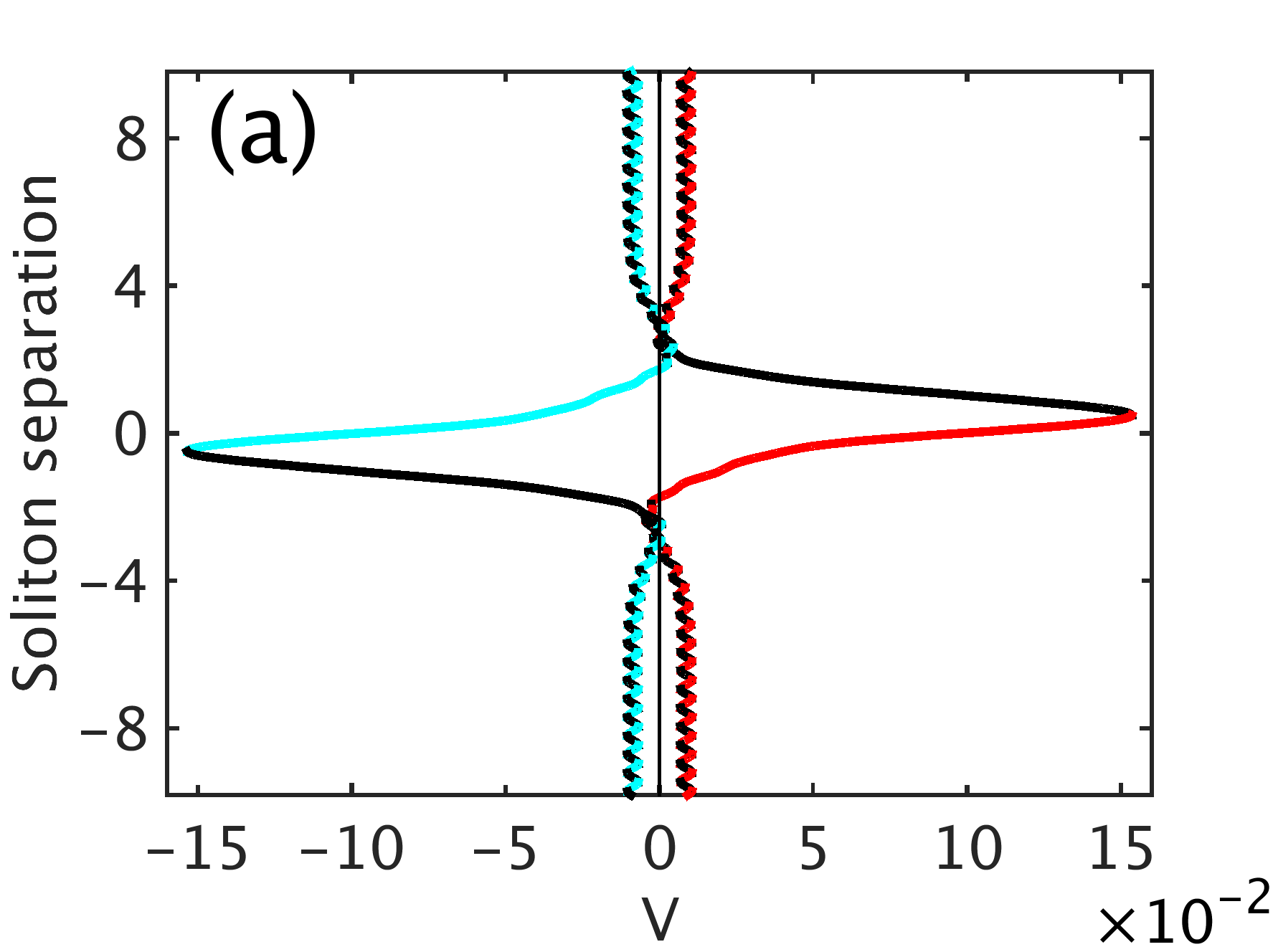}
  \end{minipage}
  \hfill
  \begin{minipage}{0.48\textwidth}
    \centering
    \includegraphics[width=\linewidth]{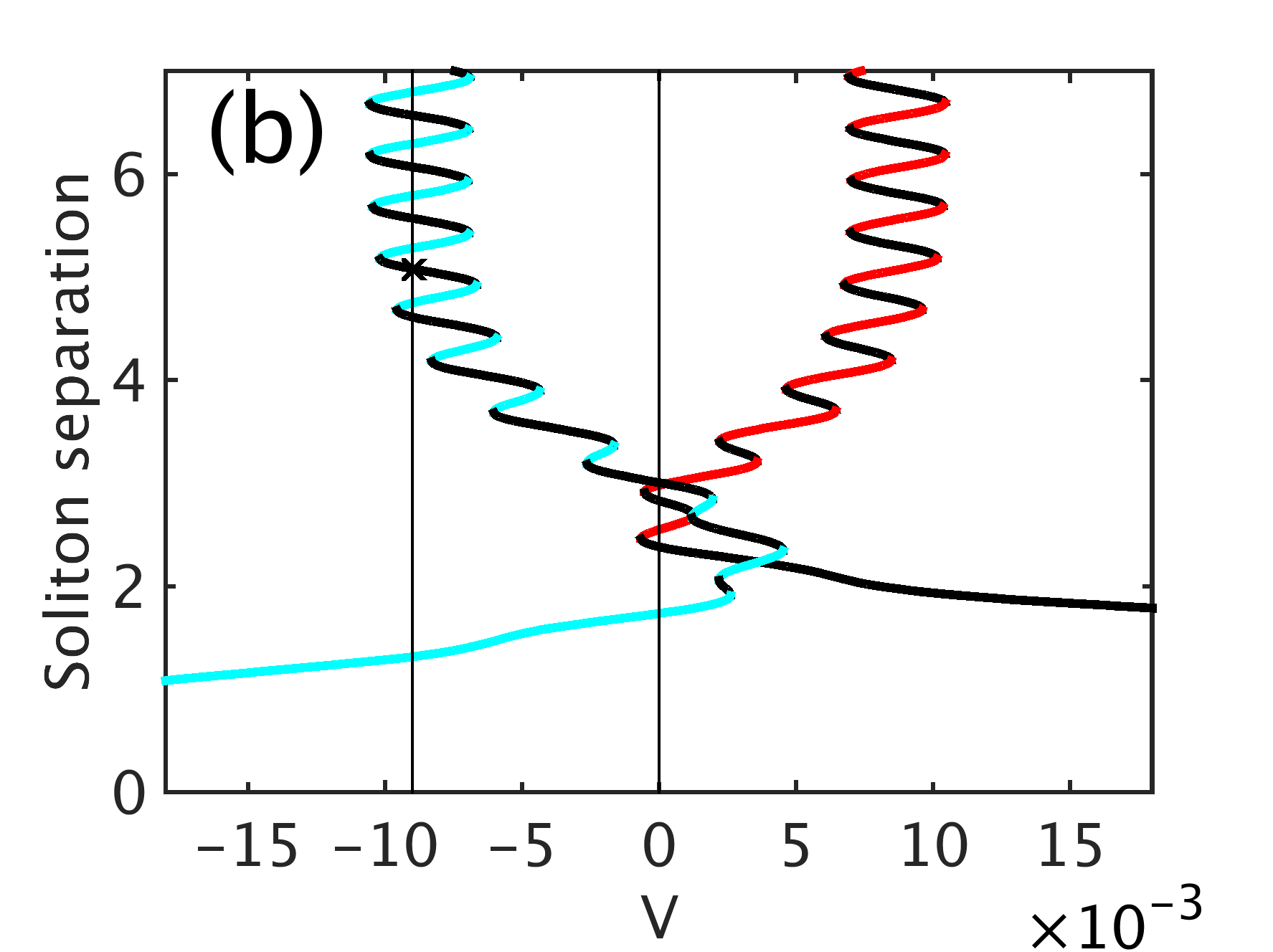}
  \end{minipage}
\caption{Soliton separation $\xi_-$ as a function of the group-velocity mismatch $V$ for the cluster branch, obtained by numerical integration of Eqs.~(\ref{eq:LLE_coupled1}) and (\ref{eq:LLE_coupled2}) using a path-following algorithm (a). Black and cyan (red) indicate the stable and unstable portions of the soliton branch for $\delta=\pi/2$ ($\delta=-\pi/2$), respectively. Panel (b): magnified view of the vicinity of the $V=0$ axis. Vertical lines indicate $V=-0.009$ and $V=0$. The black cross indicates the soliton separation in the cluster shown in Fig.~\ref{fig:map05}. All other parameters are as in Fig.~\ref{fig:delta0}.}
  \label{fig:distance05}
\end{figure}

\begin{figure}[htbp]
    \centering
    \begin{minipage}[t]{0.48\textwidth}
        \centering
        \includegraphics[width=\linewidth]{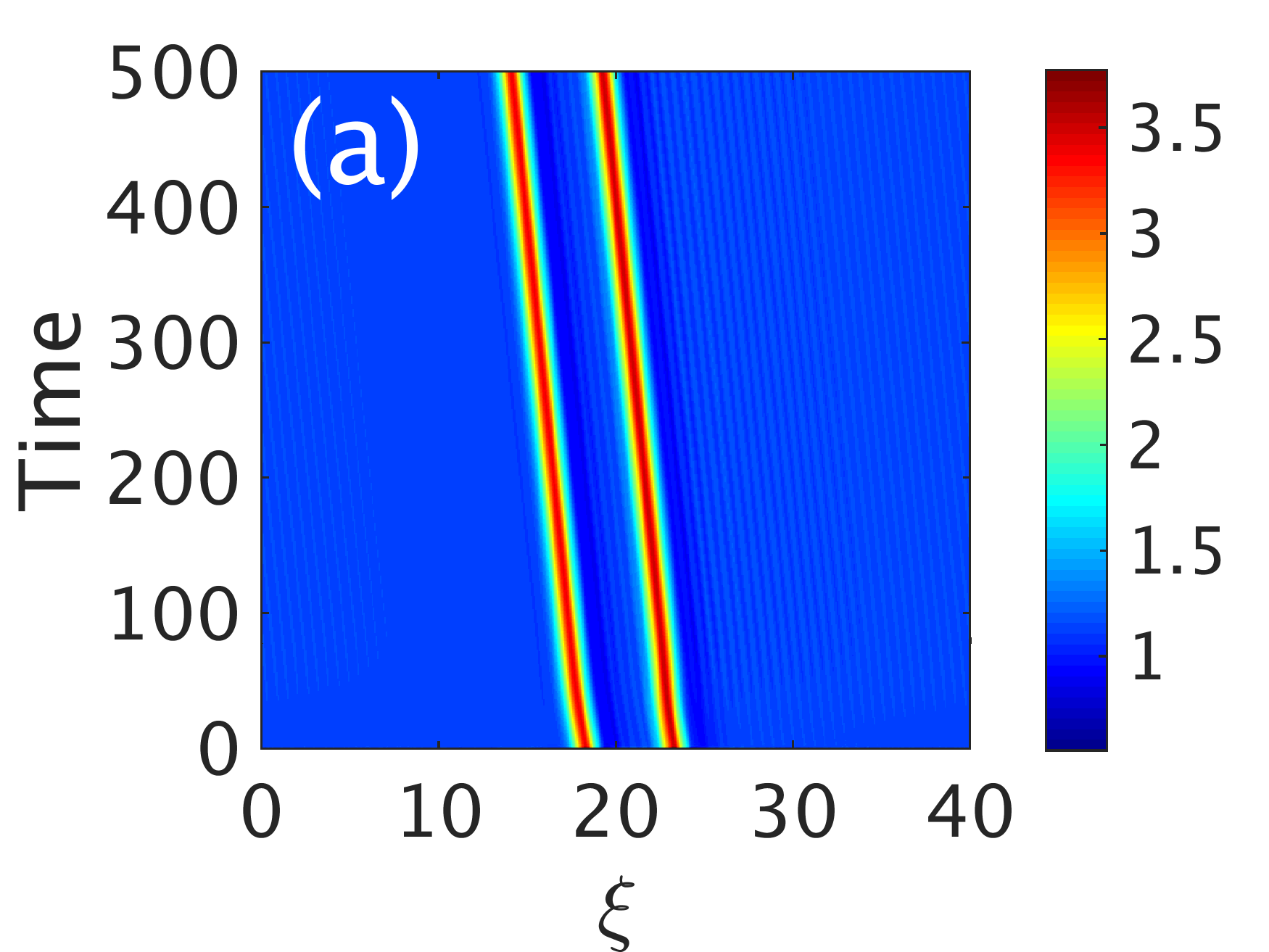}
    \end{minipage}
    \hfill
    \begin{minipage}[t]{0.48\textwidth}
        \centering
        \includegraphics[width=\linewidth]{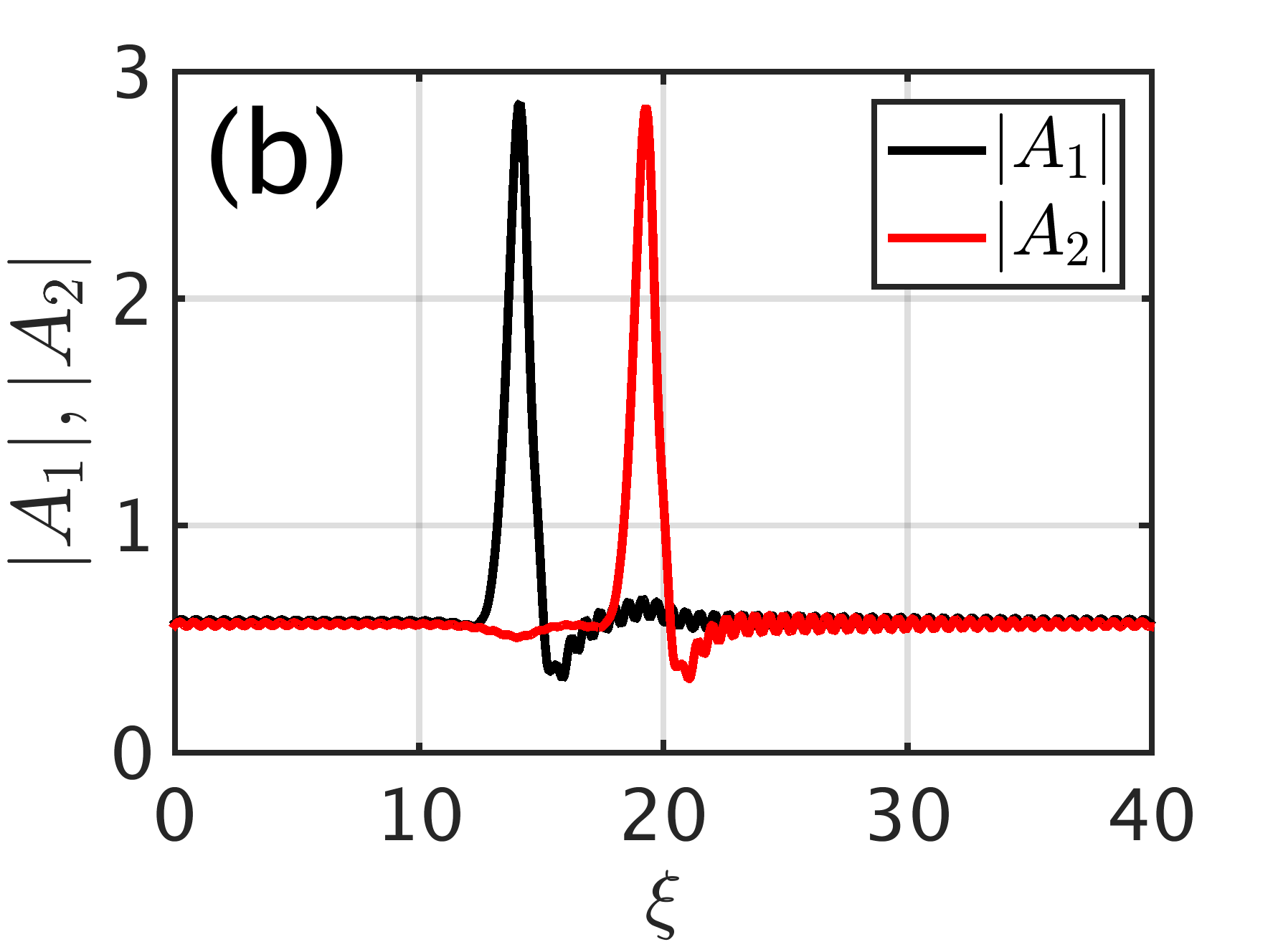}
    \end{minipage}
\caption{Soliton cluster obtained by numerical integration of Eqs.~(\ref{eq:LLE_coupled1}) and (\ref{eq:LLE_coupled2}) with $\delta=\pi/2$. This cluster corresponds to the black cross in Fig.~\ref{fig:distance05}(b).   $V=-0.009$; all other parameters are as in Fig.~\ref{fig:distance0}.}
  \label{fig:map05}
\end{figure}

\begin{figure}[htbp]
  \centering
  \begin{minipage}{0.48\textwidth}
    \centering
   \includegraphics[width=\linewidth]{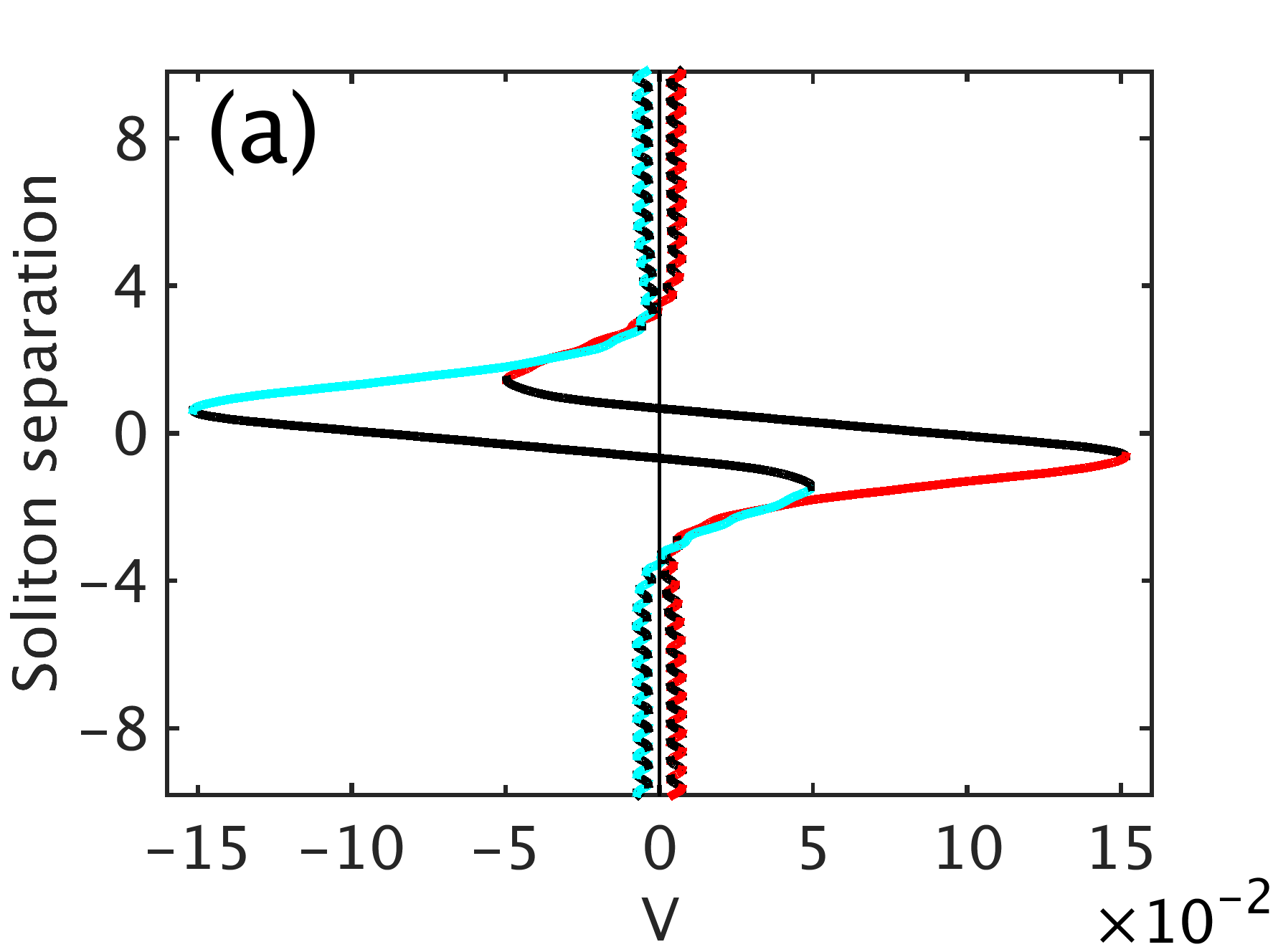}
  \end{minipage}
  \hfill
  \begin{minipage}{0.48\textwidth}
    \centering
    \includegraphics[width=\linewidth]{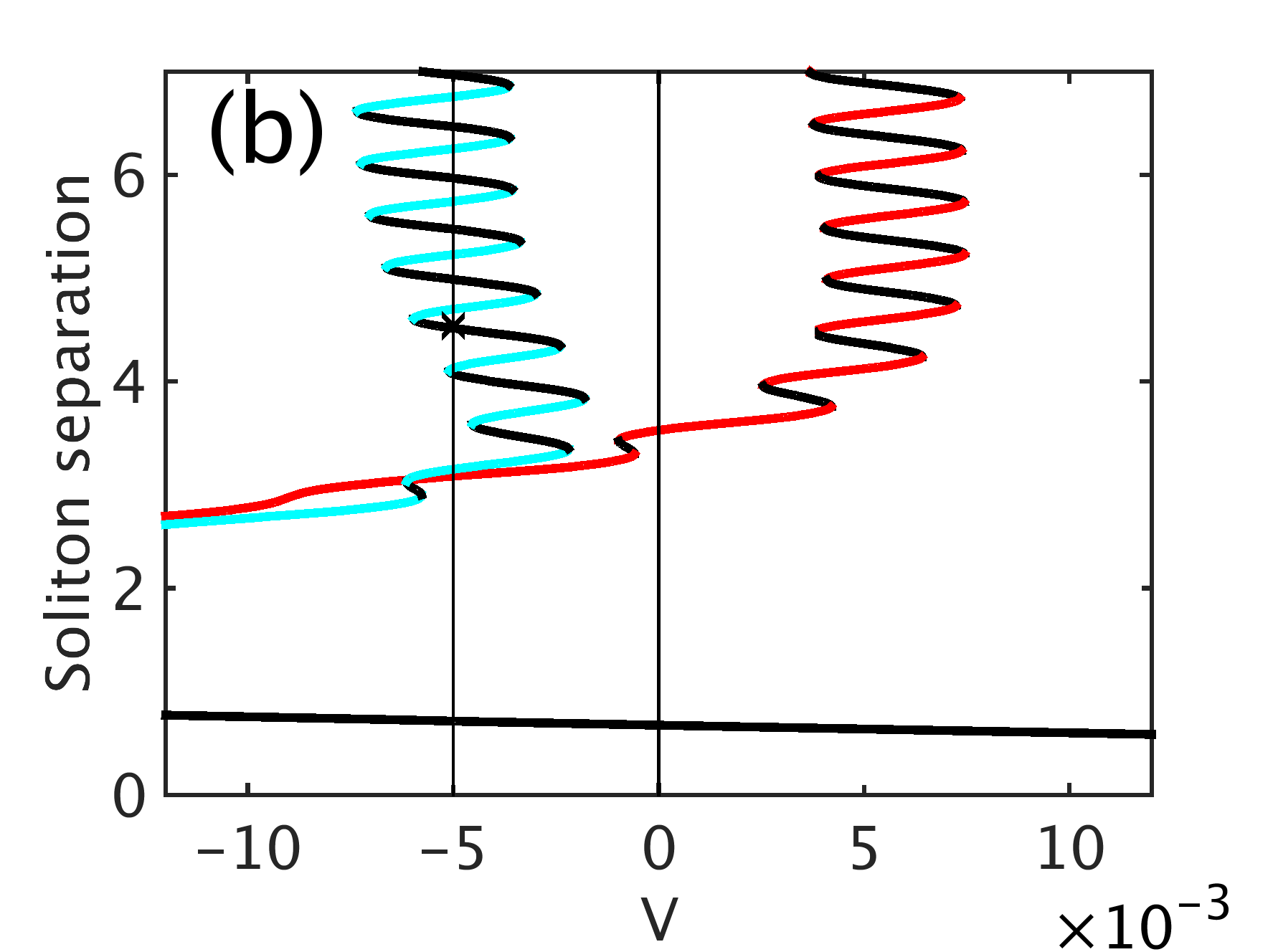}
  \end{minipage}
\caption{Soliton separation $\xi_-$ as a function of the group-velocity mismatch $V$ for the cluster branch, obtained by numerical integration of Eqs.~(\ref{eq:LLE_coupled1}) and (\ref{eq:LLE_coupled2}) using a path-following algorithm (a). Black and cyan (red) indicate the stable and unstable portions of the soliton branch for $\delta=\pi/4$ ($\delta=-\pi/4$), respectively. Panel (b): magnified view of the vicinity of the $V=0$ axis. Vertical lines indicate $V=-0.005$ and $V=0$. Here, $\kappa=0.1$; all other parameters are as in Fig.~\ref{fig:delta0}.}
  \label{fig:distance025}
\end{figure}

Figure~\ref{fig:distance05} is analogous to Fig.~\ref{fig:distance0}, but corresponds to $\delta=\pm\pi/2$.  Furthermore, owing to the property $S_-(-\xi_-)=S_-(\xi_-)$ and $C_-(-\xi_-)=-C_-(\xi)$  the interacion equation (\ref{eq:int2_TOD}) with $F_-=0$ is invariant under the transformation $(u,\xi_-,\delta)\to(-u,-\xi_-,-\delta)$. This symmetry corresponds to the transformation $(V,\delta,A_1,A_2)\to(-V,-\delta,A_2,A_1)$ of the coupled LLEs (\ref{eq:LLE_coupled1}) and (\ref{eq:LLE_coupled2}) with identical parameters. Consequently, the curve corresponding to $\delta=-\pi/2$ is obtained from that for $\delta=\pi/2$ through the transformation $(V,\xi_-)\to(-V,-\xi_-)$. This is equivalent to reversing the sign of $V$ and $\delta$ while interchanging the positions of solitons 1 and 2.

For $\delta=\pm\pi/2$, multiple-soliton cluster states can be obtained by choosing an appropriate nonzero value of the group-velocity mismatch parameter $V$. This value can be estimated from the asymptotic analysis presented in the previous section. Indeed, according to Eq.~(\ref{eq:int2_TOD}), stationary clusters for $\delta=\pm\pi/2$ are determined by the condition $u/K=\pm S_-$. As can be seen from the plot of $S_-$ in Fig.~\ref{fig:SCm}(b), multiple solutions of this equation exist, for example, at $u/K=-0.09$. Since $V/\kappa=u/K$ and $\kappa=0.1$ in Fig.~\ref{fig:distance05}, this implies that multiple-soliton clusters should exist at $V=-0.009$ for $\delta=\pi/2$. where the vertical line at $V=-0.009$ intersects the solution branch corresponding to $\delta=\pi/2$ at multiple points. The marker in this figure indicates the inter-soliton separation of the stable cluster illustrated in Fig.~\ref{fig:map05}.

\begin{figure}[htbp]
    \centering
    \begin{minipage}[t]{0.48\textwidth}
        \centering
        \includegraphics[width=\linewidth]{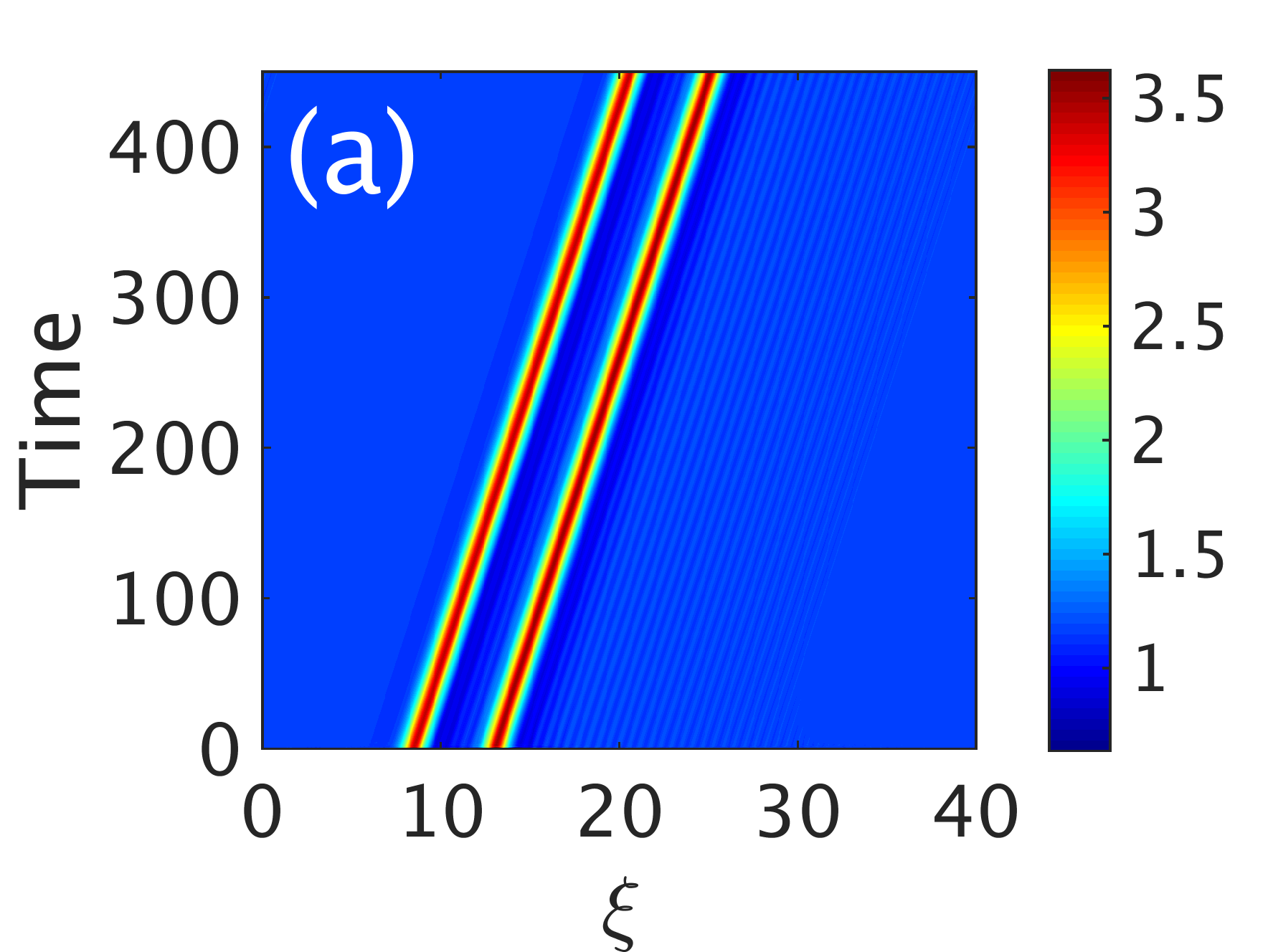}
    \end{minipage}
    \hfill
    \begin{minipage}[t]{0.48\textwidth}
        \centering
        \includegraphics[width=\linewidth]{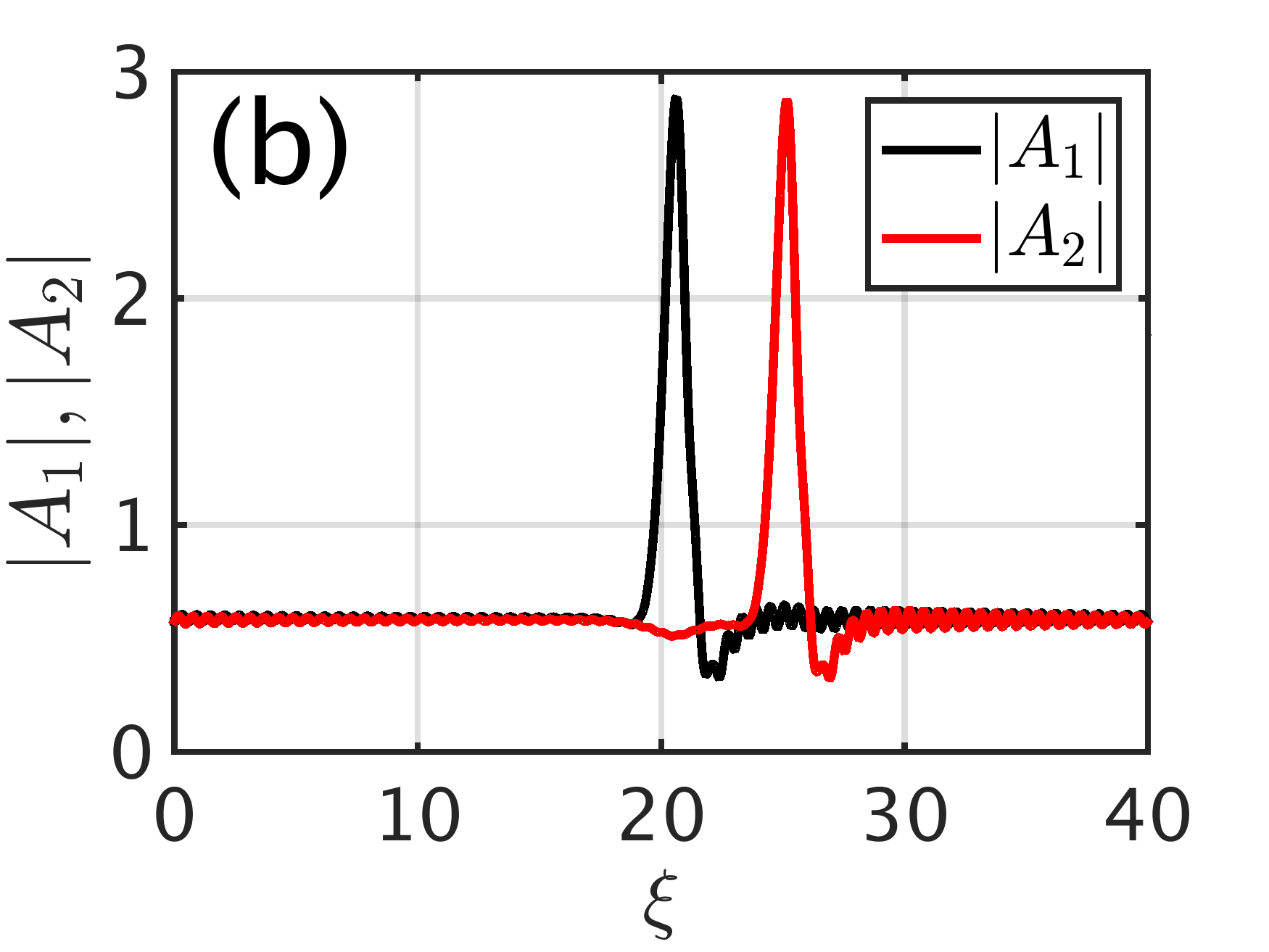}
    \end{minipage}
\caption{Soliton cluster obtained by numerical integration of Eqs.~(\ref{eq:LLE_coupled1}) and (\ref{eq:LLE_coupled2}) with $\delta=\pi/4$. This cluster corresponds to the black cross in Fig.~\ref{fig:distance025}. $V=-0.005$; all other parameters are as in Fig.~\ref{fig:distance0}.}
  \label{fig:map025}
\end{figure}

Figures \ref{fig:distance025} and \ref{fig:map025} are analogous to Figs.~\ref{fig:distance05} and \ref{fig:map05}, but correspond to $\delta=\pm\pi/4$. As in the case of $\delta=\pm\pi/2$, the branch corresponding to $\delta=\pi/4$ can be obtained from that for $\delta=-\pi/4$ through the transformation $(V,\xi)\to(-V,-\xi)$. Here multiple soliton clusters can also occur for appropriately chosen values of the group-velocity mismatch $V$. In particular, as for $\delta=\pm\pi/2$, negative (positive) values of $V$ are required for the existence of multiple-cluster states at $\delta=\pi/4$ ($\delta=-\pi/4$). However, the absolute value of $V$ must be smaller than in the case $\delta=\pm\pi/2$. This is consistent with Fig.~\ref{fig:SCm}, where the shift of the blue curve required to produce multiple intersections with the $x$ axis is smaller than the corresponding shift in Fig.~\ref{fig:SCpm}(b). In particular, the soliton cluster shown in Fig.~\ref{fig:map025} was calculated for $V=-0.005$ and corresponds to the black cross in Fig.~\ref{fig:distance025}(b).

\section{Conclusion}\raggedbottom
We have investigated the interaction of dissipative Kerr solitons in two coherently driven, evanescently coupled nearly identical Kerr microresonators in the presence TOD. Starting from coupled LLEs, we derived an asymptotic description of the slow dynamics of two interacting solitons in the regime of weak intercavity coupling and small group-velocity mismatch. The resulting equations govern the dynamics of the soliton separation and the common drift and provide an effective description of the interaction landscape in terms of the soliton profile and the corresponding adjoint translational mode.

Similar to the case of two-soliton interactions in a single Kerr cavity, TOD qualitatively modifies the soliton interaction even when its magnitude is small compared with that of the second-order dispersion. TOD breaks the parity symmetry of both the individual soliton and its adjoint neutral mode, introducing additional contributions to the interaction functions and inducing a drift of the soliton pair. As a result, the interaction between solitons becomes intrinsically asymmetric, and the structure of the resulting bound states differs qualitatively from that of the TOD-free system. The fast oscillations of one of the soliton tails, associated with Cherenkov radiation, give rise to an oscillatory interaction landscape that supports multiple stable and unstable cluster states with distinct soliton separations.

We have demonstrated that the relative phase of the coherent driving fields provides an effective control parameter for determining the type and stability of soliton clusters. For configurations sufficiently close to in-phase (antiphase) pumping of two nearly identical cavities, stable (unstable) ``bunching'' clusters and multiple bound states can coexist even in the absence of group-velocity mismatch. By contrast, for phase differences significantly removed from the in-phase and antiphase configurations, such as $\delta=\pm\pi/2$ and $\delta=\pm\pi/4$, a suitable group-velocity mismatch is required to generate multiple cluster states that do not exist at zero mismatch. The asymptotic theory provides a simple criterion for their emergence and predicts the mismatch values at which the corresponding multiple cluster branches appear.

Numerical continuation further reveals that the cluster branches terminate through successive saddle-node bifurcations as the group-velocity mismatch is varied. Beyond the final bifurcation, the solitons no longer form stationary bound states but instead propagate with a nonzero relative velocity, leading to periodic collisions. These collisions are accompanied by position shifts, indicating that the residual soliton interaction remains dynamically significant near the bifurcation point, even after the stationary cluster states have disappeared. 

The agreement between the asymptotic interaction equations and direct numerical simulations of the full coupled LLEs confirms the validity of the reduced description and shows that the essential features of soliton-cluster dynamics are captured by the reduced interaction equations. Overall, the results identify TOD, the relative pump phase, and the group-velocity mismatch as experimentally relevant control parameters for engineering dissipative-soliton arrangements in coupled microresonator systems. These findings provide a basis for the design of reconfigurable soliton clusters and coherent photonic structures in coupled-resonator platforms.

\begin{acknowledgments}
This work was supported by the RSF project No.~25-11-20069.
\end{acknowledgments}

\bibliographystyle{apsrev4-2}
\bibliography{references}

\end{document}